\documentclass[letterpaper]{article} 
\usepackage{aaai2026}  
\usepackage{times}  
\usepackage{helvet}  
\usepackage{courier}  
\usepackage[hyphens]{url}  
\usepackage{graphicx} 
\usepackage{subcaption}
\usepackage{booktabs}
\usepackage{natbib}  
\usepackage{caption} 
\usepackage[utf8]{inputenc}
\usepackage[T1]{fontenc}
\usepackage{xcolor}
\usepackage{enumitem}
\usepackage{adjustbox}
\definecolor{codebg}{gray}{0.95}
\usepackage{float}
\usepackage[linesnumbered,ruled,vlined]{algorithm2e}
\usepackage{amsmath}
\usepackage{amssymb}
\usepackage{mathtools}
\usepackage{multirow, makecell}
\usepackage{graphicx, bm}
\usepackage{amsthm}
\usepackage{tabularx}
\usepackage{array}
\usepackage{dsfont}
\usepackage{url}
\usepackage{graphicx}
\usepackage{mdframed}

\usepackage{newfloat}
\usepackage{listings}
\DeclareCaptionStyle{ruled}{labelfont=normalfont,labelsep=colon,strut=off} 
\floatstyle{ruled}
\newfloat{listing}{tb}{lst}{}
\floatname{listing}{Listing}
\title{HES-SQL: Hybrid Reasoning for Efficient Text-to-SQL with Structural Skeleton Guidance}
\author{
	Suming Qiu\equalcontrib,
	Jing Li\equalcontrib,
	Zhicheng Zhou, Junjie Huang, Linyuan Qiu, Zhijie Sun\thanks{Corresponding author.}
}

\nocopyright 

\affiliations{
    \textsuperscript{\rm 1}GTS, Huawei Technologies Co., Ltd\\
    Correspondence to: sunzhijie3@huawei.com

}

\usepackage{bibentry}

\begin{document}

\maketitle

\begin{abstract}
	We present HES-SQL, a novel hybrid training framework that advances Text-to-SQL generation through the integration of thinking-mode-fused supervised fine-tuning (SFT) with Group Relative Policy Optimization (GRPO). Our approach introduces three key innovations: (1) a skeleton-completeness scoring mechanism that enhances preference alignment between generated queries and optimal SQL structures; (2) a query-latency-aware reward system that incentivizes the generation of computationally efficient SQL queries; (3) a self-distillation process for thinking-mode completion that prevents degradation of the model's reasoning capabilities. This framework enables hybrid thinking models to switch between reasoning and non-reasoning modes while improving SQL query accuracy and execution efficiency. 
	
	Experimental evaluation, conducted on MySQL 8.0 and SQLite 3.42 under controlled single-user conditions, demonstrates that HES-SQL achieves competitive performance with execution accuracies of 79.14\% and 54.9\% on the BIRD and KaggleDBQA benchmarks, respectively. Query latency is measured as the end-to-end execution time of generated queries on the DBMS, averaged over multiple runs to mitigate variance. Efficiency gains range from 11\% to 20\% relative to supervised baselines. Our results establish a new paradigm for Text-to-SQL systems that effectively balances semantic accuracy with computational efficiency through execution-informed reinforcement learning (RL). The proposed methodology has significant implications for developing robust natural language interfaces to databases and can be extended to broader structured generation tasks requiring both correctness and efficiency optimization.
\end{abstract}

\section{Introduction}

Text-to-SQL has emerged as an important research domain in natural language processing (NLP), enabling users to interact with relational databases through natural language queries rather than complex SQL syntax. This technology helps bridge the gap between human language understanding and structured data access, making database interactions more accessible to non-technical users \citep{zhu2024largelanguagemodelenhanced}. By translating everyday language into formal queries, NL2SQL empowers a broader range of users to retrieve and analyze data, which has driven significant interest from both academia and industry. 

With the rapid advancement of large language models (LLMs), the field has witnessed significant transformations in both methodological approaches and performance benchmarks \citep{gao2023texttosqlempoweredbylargelanguage, Li_2024}. LLM-based models have substantially improved the ability to parse complex natural-language questions into SQL, achieving competitive results on benchmarks. However, developing a Text-to-SQL system that is both highly accurate \emph{and} efficient remains challenging. The complexity stems from multiple sources: understanding nuanced or ambiguous user intents, comprehending schemas with numerous tables and relations, and formulating SQL queries that not only capture the correct semantics but also execute efficiently on large-scale databases. Despite these advances, recent analyses of Text-to-SQL capabilities \citep{qin2022survey, Li_2024} emphasize that significant gaps remain in bridging robust natural language understanding with reliable SQL generation for real-world scenarios.

Recent developments in Text-to-SQL have demonstrated promising approaches. Traditional BERT-based methods, when enhanced with table content understanding, have improved execution accuracy \citep{xusheng2023research}. LLM-based solutions have further advanced the field, with several strategies achieving high execution accuracies on benchmarks through prompt engineering and fine-tuning \citep{pourreza2024dtssqldecomposedtexttosqlsmall}. For example, DAIL-SQL and related evaluations showed that carefully designed prompts and in-context examples can elicit strong performance from general-purpose LLMs \citep{gao2023texttosqlempoweredbylargelanguage}. SQLong \citep{hoang2025sqlong} introduces a data augmentation framework that synthetically extends database schemas to generate long-context training data, effectively improving LLM performance on Text-to-SQL tasks involving complex database environments. ZeroNL2SQL \citep{fan2024combining} presents a hybrid framework combining fine-tuned small language models for SQL structure identification with LLMs for semantic reasoning, achieving superior zero-shot performance across diverse environments.

Despite these advances, several limitations persist. Supervised fine-tuning (SFT) alone often faces challenges with complex reasoning and compositional generalization \citep{zhai2025excotoptimizingreasoningfortexttosql}. SFT models tend to follow surface patterns and may exhibit limited performance on queries requiring multi-step logical inference or precise condition handling, as demonstrated by Zhai et al. \citep{zhai2025excotoptimizingreasoningfortexttosql}, who incorporated execution feedback to mitigate this. The reliance on publicly deployed LLMs (e.g., GPT-4 via APIs) raises concerns about data privacy, cost, and accessibility \citep{pourreza2024dtssqldecomposedtexttosqlsmall}. Moreover, reinforcement learning (RL) approaches for Text-to-SQL are often affected by reward sparsity and instability, making optimization difficult \citep{pourreza2025reasoningsqlreinforcementlearningwithsql}. A simple reward that only distinguishes perfect SQL completions provides little signal for partial correctness, hindering the learning process. Recent studies have introduced intermediate or partial rewards to alleviate this issue \citep{pourreza2025reasoningsqlreinforcementlearningwithsql, zhang2025rewardsqlboostingtexttosqlviastepwise}, yet balancing these signals remains non-trivial. In addition, as Text-to-SQL systems move toward real-world deployments on large databases, \emph{execution efficiency} becomes critical — generated queries that are slow or overly complex may pose practical challenges. Emerging evaluations \citep{li2023can} now include metrics for execution cost (e.g., runtime or cost-based efficiency), underscoring the need for methods that optimize both correctness and performance. These challenges highlight the need for more sophisticated training methods that jointly balance semantic accuracy, computational efficiency, and practical deployment considerations.

\begin{figure}[t]
	\centering
	\includegraphics[width=0.30\textwidth]{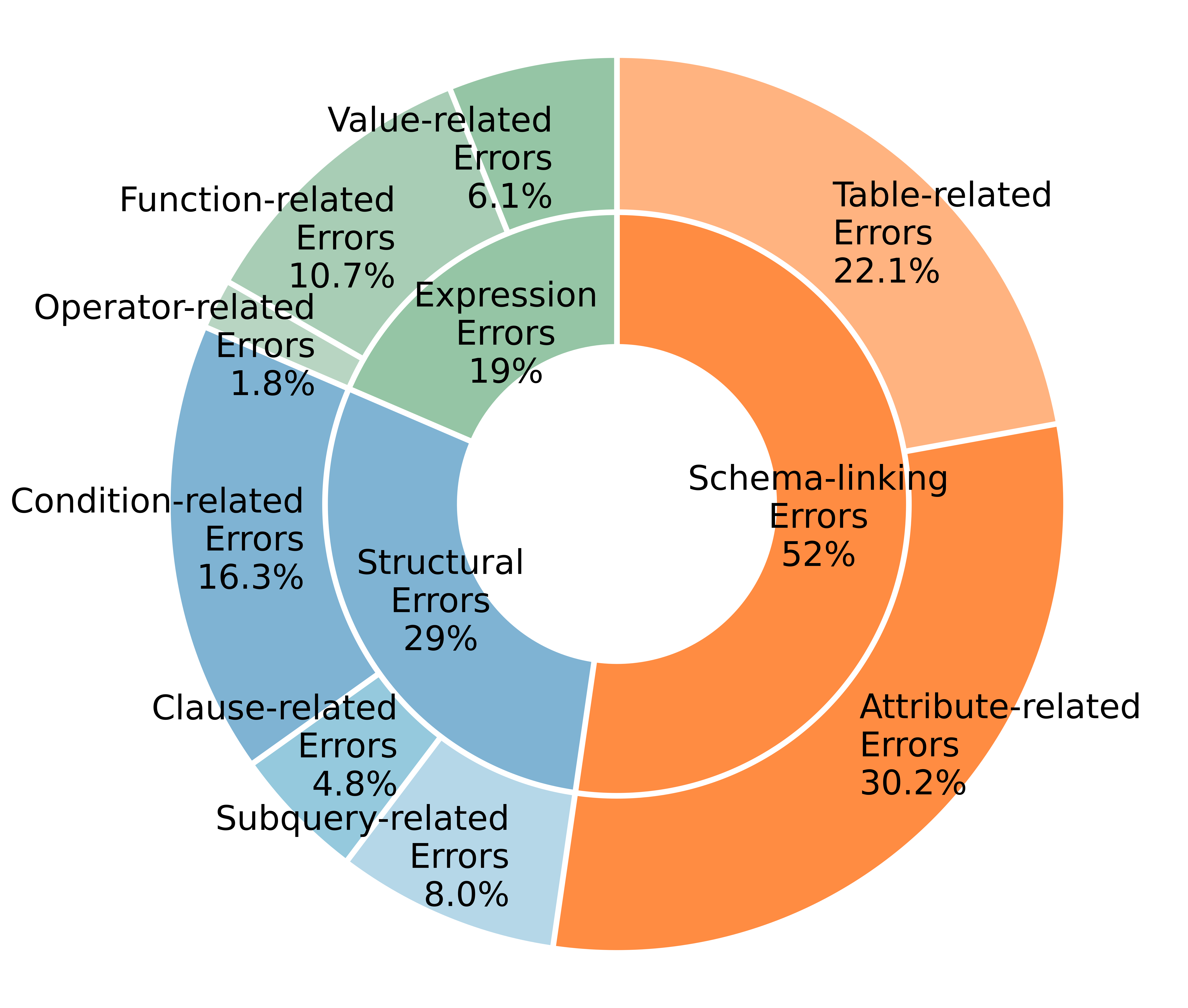}
	\caption{The error categories in BIRD benchmark using the Qwen3-32B model.}
	\label{category}
\end{figure}

Error analysis on the BIRD benchmark (Figure \ref{category}) reveals specific deficiencies in schema understanding and structural query generation. Schema-linking errors—primarily mistakes in table or attribute references—constitute over 50\% of error cases, indicating that models often struggle to accurately align natural language conditions with the correct database columns or tables. Structural errors (29\%) and condition-related errors (16\%) further highlight challenges in constructing syntactically correct SQL, such as missing subqueries or incorrect logical operators. These findings underscore the need for training methods that capture semantic–schema alignment while ensuring robust query structure generation, especially for complex queries with multiple joins or nested subqueries.

To address these issues, we present \textbf{HES-SQL}, a new hybrid training framework that combines self-distillation initialization with Grouped Relative Policy Optimization (GRPO)-based RL fine-tuning. The design of HES-SQL explicitly targets both reasoning accuracy and execution performance by introducing new reward signals and a two-stage training strategy. In essence, HES-SQL enables the model to switch between "fast" and "slow" thinking modes (non-reasoning vs. reasoning with chain-of-thought) during generation, while aligning its outputs with both semantic correctness and efficiency. Our key contributions are:

\begin{itemize}
	\item \textbf{Skeleton-completeness reward}: We propose a skeleton-completeness scoring mechanism that evaluates the structural alignment of a generated query against the ground-truth SQL pattern. This reward is used to filter candidate SQL queries, helping ensure that only syntactically and logically plausible candidates proceed to execution evaluation.
	
	\item \textbf{Latency-aware optimization}: We introduce a query-latency-aware reward mechanism that explicitly incentivizes the model to produce more efficient SQL formulations while maintaining correctness. Latency is measured as the end-to-end execution time of the generated query on the target DBMS (MySQL~8.0 and SQLite~3.42) under controlled single-user conditions, averaged over repeated runs to mitigate variance. By incorporating this signal into RL, we guide the model towards solutions that balance accuracy with lower execution cost, thus helping address practical efficiency concerns.
	
	\item \textbf{Self-distillation for reasoning modes}: We implement a progressive self-distillation strategy that initializes the model with high-quality reasoning examples (including step-by-step \texttt{<think>} processes). This step uses intermediate model checkpoints as lightweight teachers for thinking-mode completion, adding only modest training overhead. It helps preserve the model's chain-of-thought reasoning capability, mitigating the degradation of reasoning skills during RL fine-tuning and enabling the hybrid thinking model to seamlessly switch between reasoning and non-reasoning modes.
\end{itemize}

HES-SQL achieves substantial improvements in both query accuracy and execution efficiency. On the BIRD, Spider \citep{yu2018spider}, and KaggleDBQA \citep{lee-2021-kaggle-dbqa} benchmarks, it achieves execution accuracies of 79.14\%, 84.04\%, and 54.9\% respectively, consistently outperforming strong supervised fine-tuning baselines. At the same time, it yields efficiency gains of 11\%–20\% (relative reduction in latency) compared to those baselines, resulting in faster query execution on average. This dual optimization distinguishes HES-SQL from prior systems: for example, our approach achieved first place on the public BIRD leaderboard (as of submission) and ranks among the top entries on Spider, using an open-source 32B LLM. The ability to balance semantic correctness with execution efficiency positions HES-SQL as a promising approach for real-world database applications, where accurate answers need to be delivered under practical time and resource constraints.

\section{Motivation and Background}

\subsection{Text-to-SQL: Problem Formulation and Challenges}

Text-to-SQL has emerged as an important research domain in natural language processing, enabling users to interact with relational databases through natural language queries rather than complex SQL syntax. The task involves translating everyday language into formal database queries, empowering a broader range of users to retrieve and analyze data. However, developing a Text-to-SQL system that is both highly accurate and efficient remains challenging. The complexity stems from multiple sources: understanding nuanced or ambiguous user intents, comprehending schemas with numerous tables and relations, and formulating SQL queries that not only capture the correct semantics but also execute efficiently on large-scale databases. Recent analyses of Text-to-SQL capabilities emphasize that significant gaps remain in bridging robust natural language understanding with reliable SQL generation for real-world scenarios, where both correctness and computational efficiency are critical.

\subsection{Evolution of LLM-based Text-to-SQL Systems}

The emergence of LLM-based approaches has fundamentally transformed Text-to-SQL systems. Models like GPT-3/4 and their open-source counterparts have been applied to generate SQL from natural language with few-shot prompting or fine-tuning. DAIL-SQL \citep{gao2023texttosqlempoweredbylargelanguage} conducted a comprehensive evaluation of LLM capabilities in Text-to-SQL tasks, highlighting the importance of effective prompt engineering and in-context example selection. While these large models demonstrated impressive performance (often surpassing earlier specialized parsers), their analysis also revealed challenges in achieving both high accuracy and query efficiency. For instance, strong LLMs can still produce overly complex queries or misinterpret schema elements without careful guidance.

ZeroNL2SQL \citep{gu2023interleavingpretrainedlanguagemodelsand, fan2024combining} proposed innovative hybrid frameworks that combine pre-trained language models (PLMs) and LLMs to leverage complementary strengths: smaller fine-tuned models handle schema linking and skeleton generation, while the LLM focuses on complex reasoning. This two-tier approach achieved superior zero-shot performance across diverse databases, indicating the benefit of dividing the problem into sub-tasks handled by models of appropriate scale.

\subsection{Advanced Reasoning and Search Mechanisms}

Another line of research has focused on enhancing the reasoning capability of Text-to-SQL models through search and multi-step inference. SQL-o1 \citep{lyu2025sqlo1aselfrewardheuristicdynamic} introduced a self-reward heuristic search framework utilizing Monte Carlo Tree Search (MCTS) for structured exploration of the query space, effectively finding SQL solutions via trial-and-error guided by a reward. Similarly, Alpha-SQL \citep{li2025alphasqlzeroshottexttosqlusing} applied MCTS in a zero-shot setting, demonstrating that search strategies can compensate for the lack of training by exploring multiple candidate queries. 

CHASE-SQL \citep{pourreza2024chasesqlmultipathreasoningandpreference} employed a multi-agent paradigm and diverse candidate generation strategies, where different agent models produce various SQL candidates (paths) followed by a preference ranking to select the best output. These approaches demonstrate that sophisticated reasoning mechanisms and exploration can significantly improve SQL generation accuracy \citep{talaei2024chesscontextualharnessingefficient, lee2024mcssqlleveragingmultipleprompts}, particularly on complex queries that require understanding interactions of multiple conditions or nested subqueries.

\subsection{Schema Understanding and Domain Knowledge Enhancement}

Recent efforts have also concentrated on strengthening models' SQL domain knowledge and their ability to generalize. REWRITER \citep{ma2024aplugandplaynaturallanguagerewriter} developed a plug-and-play module for natural language query rewriting to handle ambiguous or ill-posed questions before SQL translation. RSL-SQL \citep{cao2024rslsqlrobustschemalinkingin} focused on robust schema linking by introducing auxiliary tasks and data augmentations that improve the model's understanding of database schemas, thereby improving accuracy without sacrificing efficiency. 

YORO \citep{kobayashi2024youonlyreadonceyoro} introduced an approach to internalize database content by encoding table information into the model's context (or parameters) so that each query can be answered with minimal DB access, significantly reducing inference cost. SENSE \citep{yang-etal-2024-synthesizing} explored synthesizing training data using both strong and weak LLMs to boost performance, illustrating that strategic data augmentation can enhance generalization to new schemas.

\subsection{Reinforcement Learning and Training Innovations}

Innovation in training methodology has emerged as a crucial research direction \citep{sheng2025cscsqlcorrectiveselfconsistencytexttosql, xie2025opensearchsqlenhancingtexttosqldynamic, Li_2024, shkapenyuk2025automaticmetadataextractiontexttosql}. Various works have explored reinforcement learning and other optimization strategies to refine Text-to-SQL performance. Reasoning-SQL \citep{pourreza2025reasoningsqlreinforcementlearningwithsql} and SQL-R1 \citep{ma2025sqlr1trainingnaturallanguageto} fine-tune Text-to-SQL models with RL, employing SQL-specific reward functions (such as partial execution results or semantic matching) to guide generation. 

Reward-SQL \citep{zhang2025rewardsqlboostingtexttosqlviastepwise} proposed a stepwise reasoning and reward framework, giving intermediate rewards for correct partial steps of the SQL, which helped the model learn the logical progression of query construction. These RL-based approaches show that carefully designed rewards can improve logical consistency and accuracy of generated queries. However, these methods often face challenges with reward sparsity and instability, making optimization difficult. A simple reward that only distinguishes perfect SQL completions provides little signal for partial correctness, hindering the learning process.

\subsection{The Efficiency Gap in Text-to-SQL Systems}

Beyond accuracy, a few studies have begun to consider the efficiency of generated queries. For example, EllieSQL \citep{zhu2025elliesql} introduced a complexity-aware routing system that directs queries to different model pipelines based on their estimated difficulty, thereby improving overall cost-efficiency in a multi-model deployment. This system-level approach underlines the importance of efficiency, but it does not directly optimize the content of the SQL queries themselves for runtime performance. As Text-to-SQL systems move toward real-world deployments on large databases, execution efficiency becomes critical—generated queries that are slow or overly complex may pose practical challenges. Emerging evaluations now include metrics for execution cost (e.g., runtime or cost-based efficiency), underscoring the need for methods that optimize both correctness and performance.

\section{Method}\label{method}

\begin{figure}[t]
	\centering
	\includegraphics[width=0.46\textwidth]{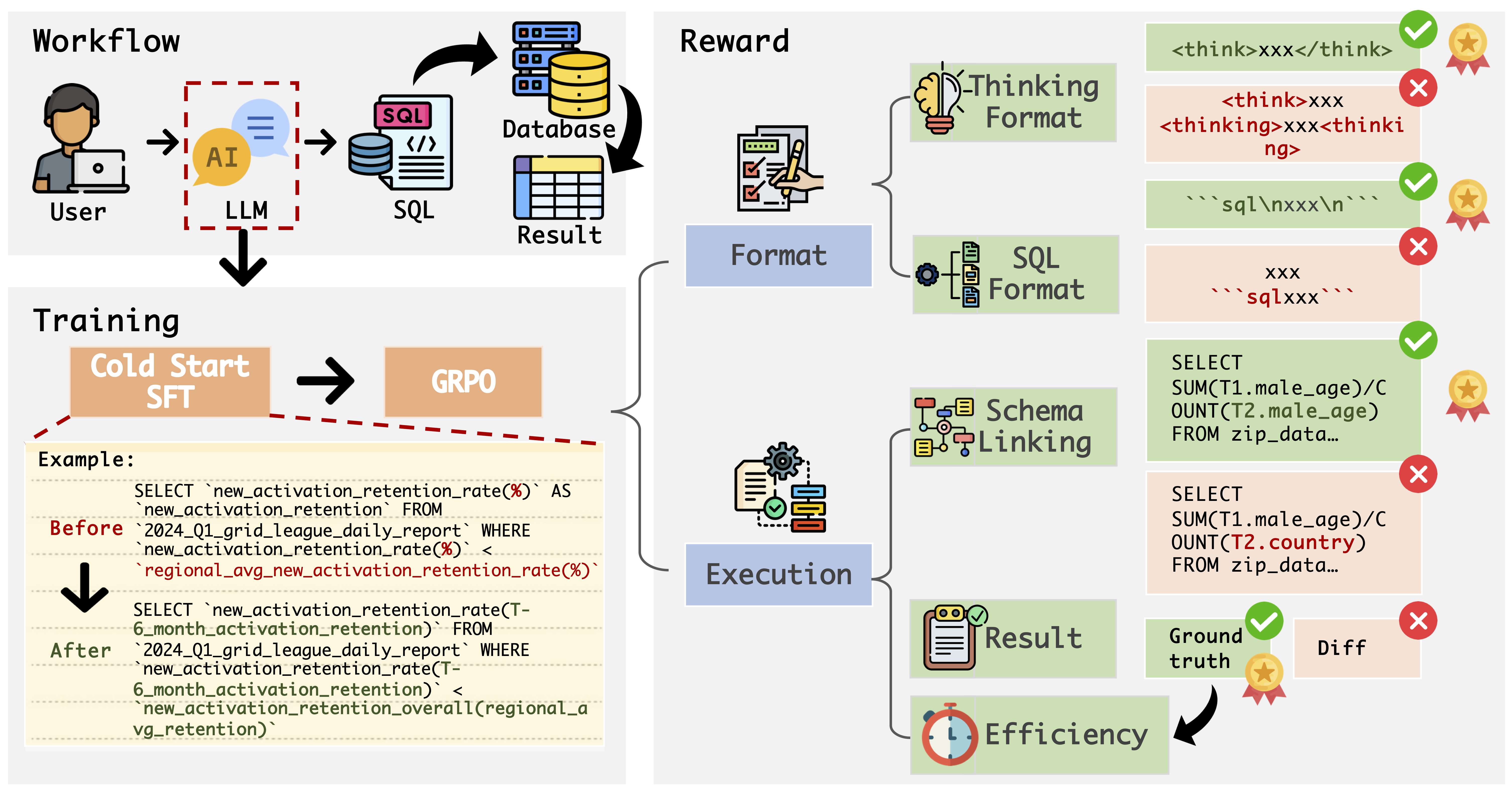}
	\caption{The architecture of the proposed hierarchical reward framework for enhanced SQL generation quality and execution efficiency.}
	\label{overall}
\end{figure}

\subsection{RL Protocol}
Policy optimization methods are the predominant paradigm for fine-tuning LLMs in Text-to-SQL tasks. In this work, we adopt GRPO as our RL strategy, leveraging its two key advantages: (1) it removes the need for auxiliary value function approximation, thereby simplifying implementation; and (2) it significantly reduces memory overhead in distributed training setups, which is critical for large-scale fine-tuning.

The policy model $\pi_{\theta}$ learns to generate structured SQL queries from natural language inputs through iterative refinement. Given an input pair consisting of a natural language question $q$ and its associated database schema $\mathcal{S}$, the model autoregressively generates a set of $G$ candidate SQL queries ${o_1, o_2, \ldots, o_G}$, where $G$ is a group size hyperparameter controlling the exploration granularity. Each candidate $o_i$ is evaluated using a composite reward function $\mathcal{R}(o_i)$ that captures the aspects including synatctic validity, execution correctness, semantic alignment, etc.

GRPO calculates relative advantages $\mathcal{A}_i$ for each candidate within a group by comparing its reward to the group mean:
\begin{equation}\label{eq:grpo_0}
	\mathcal{A}_i = \mathcal{R}(o_i) - \frac{1}{G}\sum_{j=1}^G \mathcal{R}(o_j)
\end{equation}

The GRPO objective integrates a clipped surrogate loss with KL-regularization against a reference model $\pi_{\text{ref}}$, as shown below:

\begin{equation}\label{eq:grpo_1}
	\begin{aligned}
		J_{\mathrm{GRPO}}(\theta) = \mathbb{E} 
		&\left[\frac{1}{G}\sum_{i=1}^{G}\min\left(\frac{\pi_{\theta}(o_i|q)}{\pi_{\theta_{\mathrm{old}}}(o_i|q)} \mathcal{A}_{i},\right.\right. \\
		&\left.\left.\operatorname{clip}\left(\frac{\pi_{\theta}(o_i|q)}{\pi_{\theta_{\mathrm{old}}}(o_i|q)},1-\epsilon,1+\epsilon\right) \mathcal{A}_{i}\right)\right] \\
		&-\beta D_{\mathrm{KL}}\left(\pi_{\theta} \parallel \pi_{\mathrm{ref}}\right),
	\end{aligned}
\end{equation}

Here, $\epsilon$ controls the policy update clipping range, while $\beta$ regulates the KL-penalty strength. This objective includes two essential components:

\begin{enumerate}[noitemsep]
	\item \textbf{Clipped Surrogate Objective}: The minimum of unclipped and clipped policy ratios prevents overly large updates when the ratio $\frac{\pi_{\theta}}{\pi_{\theta_{\mathrm{old}}}}$ falls outside the range $(1-\epsilon, 1+\epsilon)$.
	\item \textbf{KL-Divergence Penalty}: The regularization term $\beta D_{\mathrm{KL}}\left(\pi_{\theta} \parallel \pi_{\mathrm{ref}}\right)$ preserves linguistic capabilities learned during supervised pretraining.
\end{enumerate}

The group-based candidate generation mechanism enables robust policy learning through intra-group comparisons of structurally diverse SQL candidates, effectively handling linguistic ambiguity and the combinatorial space of semantically equivalent queries via relative advantage estimation.

\subsection{Overall Reward Design}

We designed a hierarchical reward mechanism to guide the model's SQL generation toward correctness and efficiency, as shown in Figure~\ref{overall}. The scoring system consists of four sequential stages, and the final score is their sum:

\begin{itemize}
	\item \textbf{Format Validation Stage}:  
	Each generated SQL is first checked for syntactic and structural validity (e.g., presence of \texttt{SELECT}, \texttt{FROM}). Invalid statements incur $-2$ points, while valid ones gain $+1$ point and proceed to execution. This step ensures minimal soundness and reduces downstream execution errors.
	
	\item \textbf{Execution Stage}:  
	In this stage, the syntactically valid SQL statement is executed on the target DBMS instance. Successful execution---defined as both compiling and returning results without runtime errors---earns an additional reward of $+2$ points, and the process terminates at this stage. By contrast, statements that fail execution incur a penalty of $-2.5$ points, reflecting the higher cost of execution failures compared to format violations. This design mirrors real-world deployment settings, where runtime errors are particularly detrimental.
	
	\item \textbf{Schema Linking Stage}:  
	For failed executions, we assess schema linking by checking whether referenced tables and columns match the ground-truth schema. Correct grounding earns $+1.5$ points, highlighting schema-level understanding even when execution fails.

	\item \textbf{Efficiency Stage}:  
	Executed SQL is benchmarked against the gold-standard query on MySQL~8.0 and SQLite~3.42 under identical settings. Each query is run multiple times and averaged to reduce variance. The reward is defined as $\text{time\_ratio} = \min(1.0, \tfrac{t_{gold}}{t_{pred}})$, penalizing slower queries while granting maximum reward to those matching or exceeding the gold standard.
	
\end{itemize}
The overall algorithm workflow is detailed in Appendix. In addition, the progressive self-distillation step is implemented by reusing intermediate checkpoints as lightweight teachers, introducing only modest overhead compared to baseline fine-tuning and scaling efficiently to larger model sizes. Specifically, subsequent sections will present the detailed implementation of format validation.

\subsection{Format Validation Strategy}

The format validation employs an evaluation framework aligned with dual-system cognitive theory (fast and slow thinking) modalities from Qwen3:

\begin{figure}[t]
	\centering
	\includegraphics[width=0.46\textwidth]{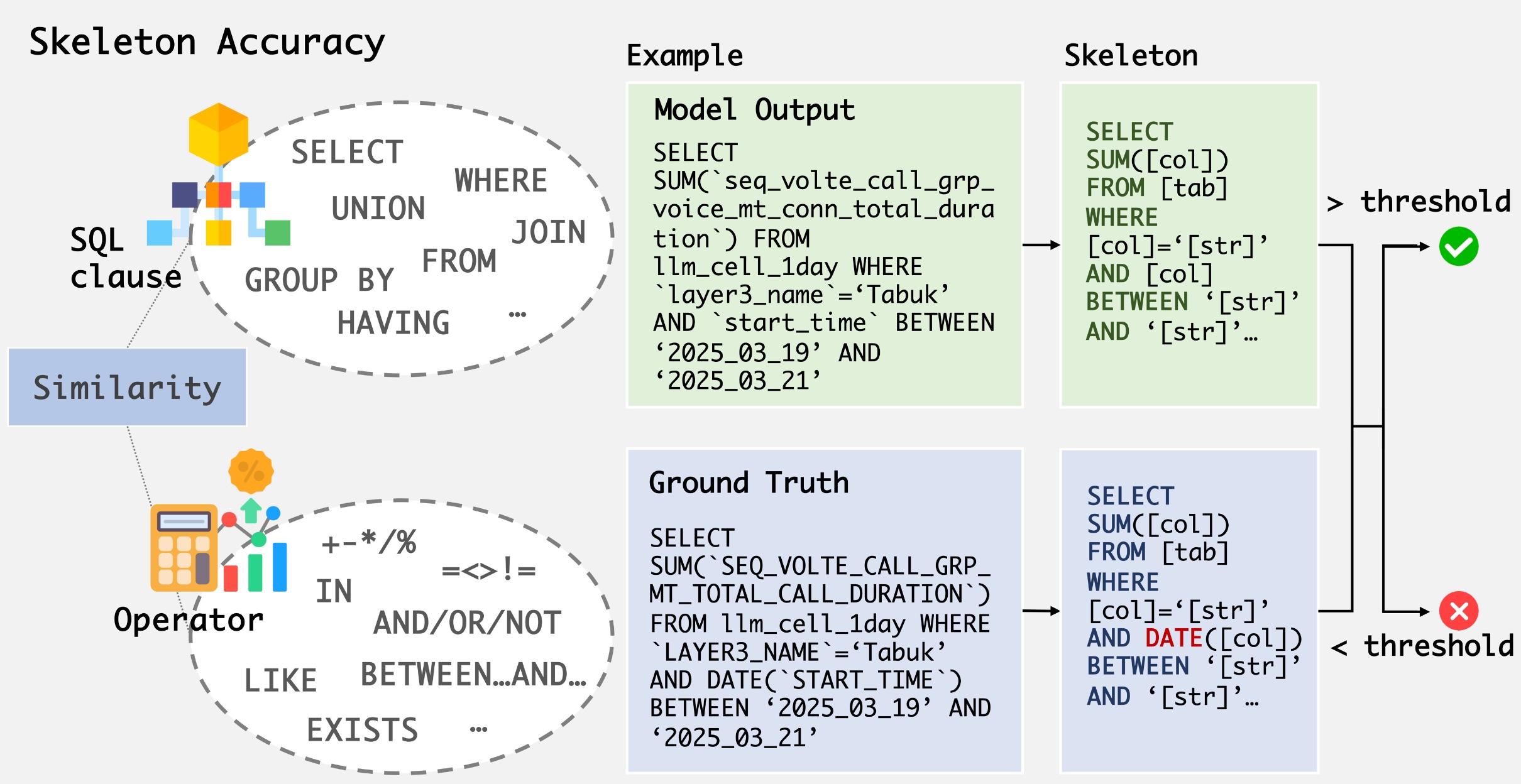}
	\caption{The skeleton accuracy mechanism adopted in reward design.}
	\label{skeleton}
\end{figure}

\begin{description}[
	leftmargin=2em,
	itemindent=-1em,
	font=\normalfont\bfseries,
	labelsep=0.5em
	]
	
	\item[Mode 1 (Supressed Thinking)] \hfill \\
	\hspace*{-1.4em}
	\fbox{\begin{minipage}[t]{0.95\columnwidth}
			\smallskip
			\textbf{Template}: \\
			\smallskip
			\texttt{user: \{query\}} $\rightarrow$ \texttt{assistant: \{answer\}}
			
			\textbf{Constraints}:
			\begin{itemize}[nosep,leftmargin=1.5em]
				\item Strictly prohibits \texttt{<think>} tags in the output
				\item Direct answer without reasoning
				\item Answer must use Markdown format \\
				(\texttt{\`{}\`{}\`{}sql\textbackslash{}n***\textbackslash{}n\`{}\`{}\`{}})
			\end{itemize}
			\smallskip
	\end{minipage}}
	
	\item[Mode 2 (Fast Thinking)] \hfill \\
	\hspace*{-1.4em}
	\fbox{\begin{minipage}[t]{0.95\columnwidth}
			\smallskip
			\textbf{Template}: \\
			\texttt{user: \{query\}/no\_think} $\rightarrow$ \texttt{assistant: <think>\textbackslash n\textbackslash n</think>\textbackslash n\textbackslash n \{answer\}}
			
			\textbf{Constraints}:
			\begin{itemize}[nosep,leftmargin=1.5em]
				\item \texttt{<think></think>} tags must exist
				\item Tags must contain no content with two line feed (\textbackslash n)
				\item SQL response wrapped in Markdown format \\
				(\texttt{\`{}\`{}\`{}sql\textbackslash{}n***\textbackslash{}n\`{}\`{}\`{}})
			\end{itemize}
			\smallskip
	\end{minipage}}
	
	\item[Mode 3 (Slow Thinking)] \hfill \\
	\hspace*{-1.4em}
	\fbox{\begin{minipage}[t]{0.95\columnwidth}
			\smallskip
			\textbf{Template}: \\
			\texttt{user: \{query\}/think} $\rightarrow$ \texttt{assistant: <think>\{reasoning\}</think>\textbackslash n\textbackslash n\{answer\}}
			
			\textbf{Constraints}:
			\begin{itemize}[nosep,leftmargin=1.5em]
				\item \texttt{<think></think>} tags must exist
				\item Non-empty reasoning content required
				\item SQL response wrapped in Markdown format \\
				(\texttt{\`{}\`{}\`{}sql\textbackslash{}n***\textbackslash{}n\`{}\`{}\`{}})
			\end{itemize}
			\smallskip
	\end{minipage}}
\end{description}

As illustrated in Figure~\ref{skeleton}, we perform normalization and keyword weighting to evaluate the structural correctness of the generated SQL compared to ground truth, with the methodology outlined below:

\begin{enumerate}
	\item \textbf{Skeleton Extraction}:
	The extraction process converts raw SQL into a structural skeleton through systematic placeholder substitution. All field names are replaced with \texttt{[col]} markers, string literals with \texttt{[str]}, and numeric values with \texttt{[val]}. This transformation is illustrated below:
	\begin{quote}
		\small\ttfamily
		\begin{tabular}{@{}l@{}}
			Original SQL: \\
			SELECT SUM(`seq\_volte\_call\_grp\_voice`) \\
			FROM llm\_cell\_1day \\
			WHERE `layer3\_name` = 'Tabuk' \\
			AND `start\_time` BETWEEN '2025-03-19' \\
			AND '2025-03-21'
		\end{tabular}
		
		\centering $\downarrow$ \\ 
		
		\begin{tabular}{@{}l@{}}
			Extracted Skeleton: \\
			SELECT SUM([col]) \\
			FROM [tab] \\
			WHERE [col] = '[str]' \\
			AND [col] BETWEEN '[str]' AND '[str]'
		\end{tabular}
	\end{quote}

	\item \textbf{Weighted Emphasis}:
	Critical SQL clauses receive weighted emphasis through intentional repetition. The \texttt{WHERE} clause keywords are tripled (e.g., \texttt{WHERE} \texttt{WHERE} \texttt{WHERE}), while \texttt{JOIN} and \texttt{GROUP BY} clauses are doubled. For instance:
	\begin{quote}
		\noindent
		\begin{minipage}{\linewidth}
			\small\ttfamily
			\begin{tabular}{@{}p{1.0\linewidth}@{}}
				SELECT A FROM B WHERE condition = C
			\end{tabular}
			
			\centering $\downarrow$ \\
			
			\begin{tabular}{@{}p{1.0\linewidth}@{}}
				SELECT A FROM B WHERE WHERE WHERE condition = C
			\end{tabular}
		\end{minipage}
	\end{quote}
	
	\item \textbf{Similarity Calculation}:
	Structural similarity evaluation employs a hybrid metric combining sequence alignment and set comparison. The final score derives from:
	\begin{equation}
		\text{Similarity} = \alpha \underbrace{\text{Match Ratio}}_{\text{difflib}} + (1 - \alpha)  \underbrace{\left(\frac{|S_{\text{gen}} \cap S_{\text{gt}}|}{|S_{\text{gen}} \cup S_{\text{gt}}|}\right)}_{\text{Jaccard}}
	\end{equation}
	Where $S_{\text{gen}}$ and $S_{\text{gt}}$ represent the token sets of generated and ground-truth skeletons respectively. The \textsc{MatchRatio} component evaluates edit-distance similarity at character level, while \textsc{Jaccard} index measures keyword presence consistency. Through experimentation, the optimal performance is achieved when $\alpha$ is set to 0.7.
\end{enumerate}

The resulting composite similarity score is compared against a predefined threshold. Statements surpassing this threshold are labeled as \texttt{pass}; otherwise, they are marked as failing \texttt{fail}.

This structured method effectively integrates cognitive modeling with precise structural evaluation, guiding the model toward more accurate and semantically meaningful SQL generation.

\section{Experiments}

\label{experiment}

\begin{table}[t]
	\centering
	\caption{Models employed in ablation studies.}
	\label{tab:ablation}
	\setlength{\extrarowheight}{2pt}
	\begin{tabular}{|p{0.2\linewidth}|p{0.7\linewidth}|}
		\hline
		Model & Description \\
		\hline
		Baseline & Qwen3 32B model \\
		\hline
		Vanilla & RL with \textbf{vanilla} reward (including reward for thinking format, SQL format, schema linking, and execution) based on \textbf{Baseline} \\
		\hline
		Skeleton & RL with \textbf{vanilla} and \textbf{skeleton} reward based on \textbf{Baseline} \\
		\hline
		Efficiency & RL with \textbf{vanilla} and \textbf{efficiency} reward based on \textbf{Baseline} \\
		\hline
		Skeleton + Efficiency & RL with \textbf{vanilla}, \textbf{skeleton}, and \textbf{efficiency} rewards based on \textbf{Baseline} \\
		\hline
		HES-SQL & After \textbf{cold-start SFT}, RL with all reward mentioned above based on \textbf{Baseline} \\
		\hline
	\end{tabular}
\end{table}

\begin{table*}[t]
	\centering
	\caption{The performance ranking of HES-SQL on BIRD, Spider, and KaggleDBQA benchmarks.}
	\label{leaderboard}
	\setlength{\tabcolsep}{1pt}
	\begin{tabular}{clr|clr|clr}
		\toprule
		Rank & Model & BIRD (EX) & Rank & Model & Spider (EX) & Rank & Model & KaggleDBQA (EM) \\
		\midrule
		1 & \textbf{HES-SQL} & \textbf{79.14} & 1 & MiniSeek & 91.20 & 1 & RAT-SQL & 26.77 \\
		2 & AskData & 75.36 & 2 & DAIL-SQL* & 86.60 & 2 & \textbf{Skeleton + Efficiency} & \textbf{21.98} \\
		3 & CHASE-SQL & 74.90 & 3 & DAIL-SQL & 86.20 & 3 & \textbf{Vanilla} & \textbf{21.00} \\
		4 & TCDataAgent-SQL & 74.12 & 4 & DIN-SQL & 85.30 & 4 & \textbf{HES-SQL} & \textbf{17.58} \\
		5 & XiYan-SQL & 73.34 & 5 & \textbf{HES-SQL} & \textbf{84.04} & 5 & Edit-SQL & 11.73 \\
		\bottomrule
	\end{tabular}
\end{table*}

\begin{table*}[t]
	\centering
	\caption{The accuracy performance comparison across Text-to-SQL benchmarks.}
	\label{accuracy}
	\begin{tabular}{l*{9}{r}}
		\toprule
		Model & \multicolumn{3}{c}{BIRD} & \multicolumn{3}{c}{Spider} & \multicolumn{3}{c}{KaggleDBQA} \\
		\cmidrule(lr){2-4} \cmidrule(lr){5-7} \cmidrule(lr){8-10}
		& EM            & PGR               & TEP              & EM              & PGR              & TEP             & EM              & PGR            & TEP \\
		\midrule
		Baseline             & 0.0424        & -                 & -                & 0.2195          & -                & -               & 0.1868          & -              & - \\
		Vanilla     & 0.0437        & \textbf{228.9231} & -5.1557          & 0.2244          & -12.0408         & 0.2290          & 0.2100            & -0.4741        & 0.2974 \\
		Skeleton  & 0.0200          & -13.2857          & -1.5149          & 0.2408          & -2.7700            & 0.3273          & 0.1209          & 0.1669         & 0.2415 \\
		Efficiency & 0.0463        & 76.3077           & 12.4175          & 0.2137          & \textbf{10.1724} & 0.1508          & 0.1913          & -2.4444        & 0.0129 \\
		Skeleton + Efficiency       & 0.1317        & 3.3326            & \textbf{-6.4448} & \textbf{0.2785} & -1.0000               & 0.5247          & \textbf{0.2198} & -0.3333        & 0.3464 \\
		HES-SQL & \textbf{0.3400} & 1.0000                 & -3.6854          & 0.1605          & 1.0000                & \textbf{0.8276} & 0.1758          & \textbf{1.0000}     & \textbf{2.0393} \\
		\bottomrule
	\end{tabular}
\end{table*}

\subsection{Experimental Setup} 

\paragraph{\textbf{Training Configuration}} The experiments utilized a distributed computing infrastructure with 16 nodes containing 8 Ascend 910 processors each, employing 32GB Ascend 910B4 processors for SFT and 64GB Ascend 910B3 processors for RL training. The distributed training implemented HCCL backend with hybrid parallelism (tensor parallel size 8, pipeline parallel size 4), global batch size 128, and sequence length 8,192 tokens. The actor model employed Adam optimizer ($\beta_1=0.9$, $\beta_2=0.95$) with constant learning rate $1\times10^{-6}$ and weight decay 0.01.

\paragraph{\textbf{Self-distillation SFT and RL Training}} 
Since the BIRD and Spider training datasets lack reasoning processes, the Qwen3-32B model is utilized to generate thinking processes and corresponding SQL statements based on training queries. During the screen process, only reasoning sequences that produce SQL outputs exactly matching the ground truth are considered valid and incorporated into the \texttt{<think></think>} tags of our SFT data, while non-matching outputs are filtered out. Appendix presents the details of training data.

We conduct ablation studies to assess the individual contributions of different reward components in Text-to-SQL enhancement using Qwen3 32B as the foundation model. Table~\ref{tab:ablation} outlines the experimental variants, systematically incorporating basic rewards (format, syntax, schema linking, execution), skeleton rewards, and efficiency rewards through RL. The final configuration explores self-distillation SFT followed by comprehensive RL training, enabling precise evaluation of each reward component's impact on model performance.

To control computational cost, the progressive self-distillation uses intermediate checkpoints as lightweight teachers and performs a single additional SFT pass on the distilled data; RL steps remain unchanged. 
This design adds only modest training overhead and no extra inference-time latency.

\paragraph{\textbf{DBMS Execution Environment}}
All latency and efficiency measurements are conducted on \textbf{MySQL~8.0} and \textbf{SQLite~3.42}. 
Unless otherwise stated, queries are executed under controlled single-user load to avoid cross-query interference. 
For each natural-language input, we run both the gold SQL and the model-predicted SQL on the same DBMS instance, perform a short warm-up, and then measure \emph{end-to-end} execution time over multiple repetitions (reporting the arithmetic mean). 
We fix the dataset snapshot, schema, and DBMS configuration throughout experiments to minimize variance. 
When aggregating efficiency across engines, we first compute per-engine metrics and then report the average (MySQL/SQLite) for each benchmark.

\subsection{Evaluation Metrics}
In real-world data analysis applications, Text-to-SQL models are primarily evaluated based on their capacity to deliver accurate and efficient query results when processing large-scale database environments. To address this evaluation requirement, we introduce several complementary metrics within existing advanced benchmarks: exact-setmatch accuracy (EM), execution accuracy (EX), valid efficiency score (VES), etc. Detailed evaluation metrics are described in Appendix.

\noindent\textbf{Efficiency Metric (VES).} 
For a successfully executed prediction, $\mathrm{time\_ratio}=\min\!\left(1.0,\frac{t_{\mathrm{gold}}}{t_{\mathrm{pred}}}\right)$ is used to compute per-query efficiency factor, 
where $t_{\mathrm{gold}}$ and $t_{\mathrm{pred}}$ denote the end-to-end execution times measured on the same DBMS (MySQL~8.0 or SQLite~3.42) under single-user load, averaged across repeated runs. 
A VES score above $1.0$ (after aggregation across queries) indicates that generated queries execute faster than the reference solutions while remaining correct. 
Unless otherwise noted, we report VES as the mean of the MySQL and SQLite results to reduce engine-specific bias.

\begin{figure}[t]
	\centering
	\includegraphics[width=0.45\textwidth]{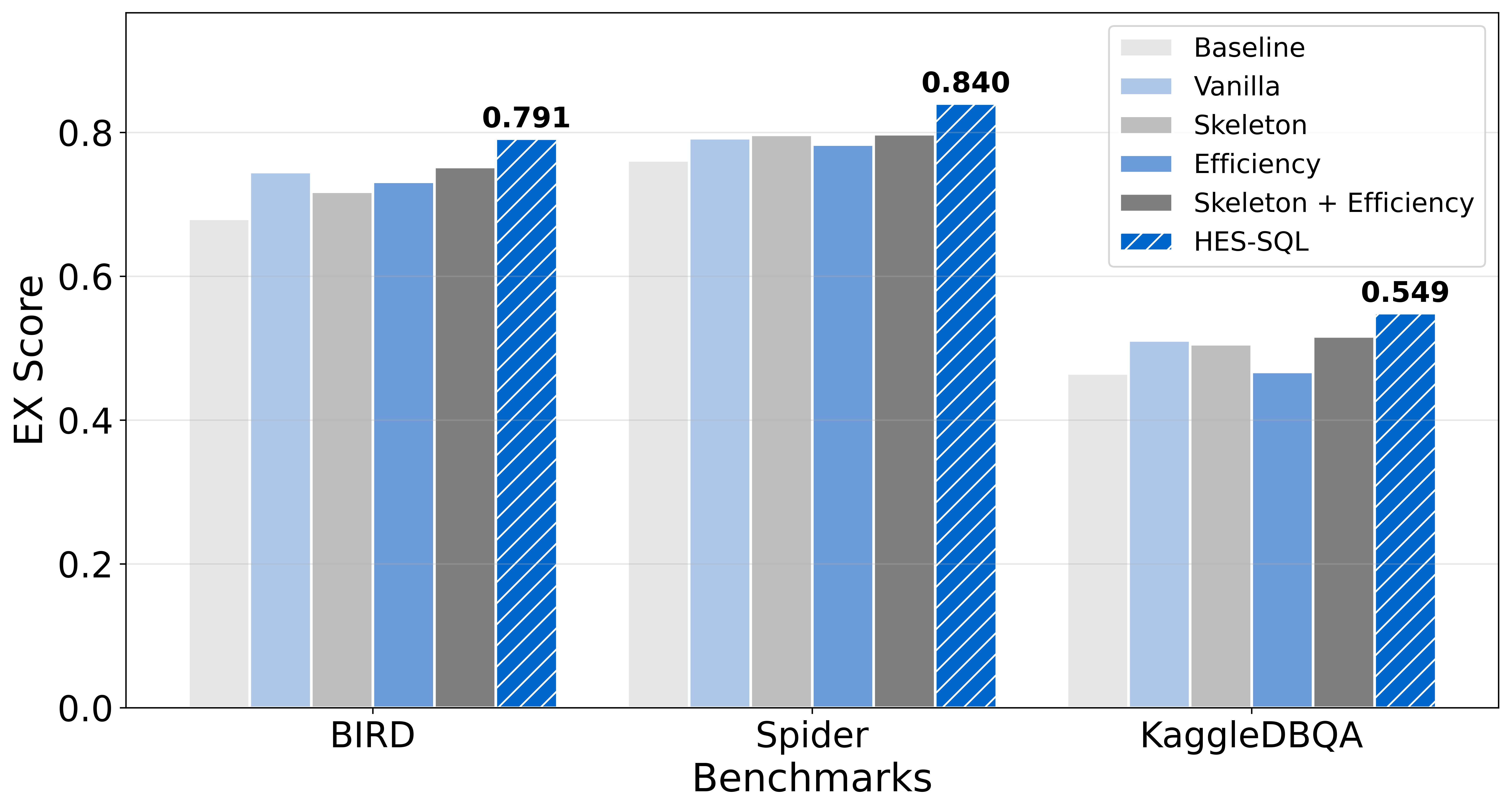}
	\caption{The execution accuracy comparison across models and Text-to-SQL benchmarks.}
	\label{overall_accuracy}
\end{figure}

\subsection{Ablation Study}

\paragraph{\textbf{Overall Evaluations on Benchmarks}}
A comparison was conducted between HES-SQL and its ablation study variants against top-performing models on the BIRD, Spider, and KaggleDBQA leaderboards, with results presented in Table~\ref{leaderboard}. HES-SQL consistently achieves competitive rankings among the top-performing methods across all three benchmarks. The training data contains only the BIRD training dataset and a minimal portion of the Spider training dataset (15\%), yet the superior performance on the KaggleDBQA test set demonstrates strong generalization capabilities.

The EX performance across BIRD, Spider, and KaggleDBQA is presented in Figure~\ref{overall_accuracy} for our ablation study. The most significant gains emerge from the HES-SQL, which performs composite training with self-distillation SFT for initialization. It consistently outperforms all other configurations across benchmarks (BIRD: +16.4\%, Spider: +10.42\%, KaggleDBQA: +18.06\% over the baseline). This two-stage training paradigm demonstrates that self-distillation SFT enables the model to learn standard SQL patterns through ground truth-consistent thinking processes, achieving high accuracy levels and providing a robust initialization for subsequent RL, which enables more effective reward signal utilization and accelerated convergence to superior performance levels.

\begin{figure}[t]
	\centering
	\includegraphics[width=0.46\textwidth]{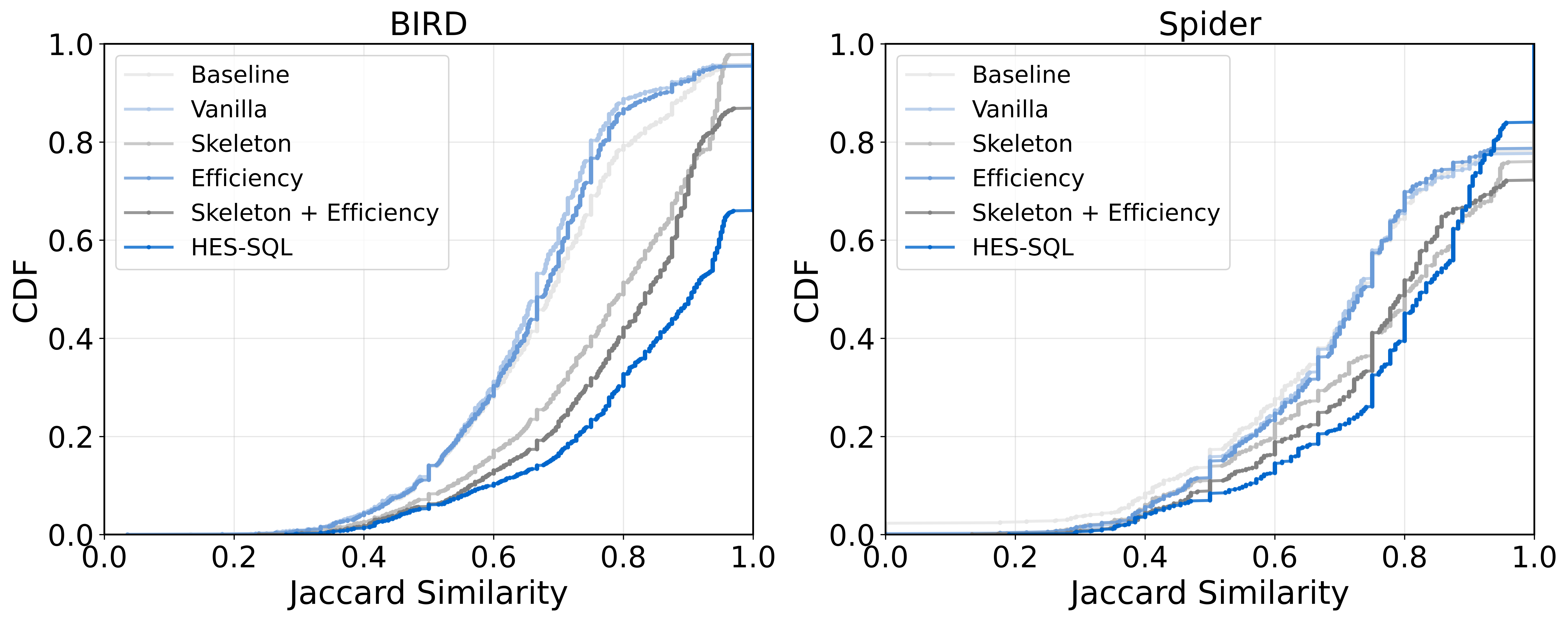}
	\caption{The CDF analysis of SQL query similarity using Jaccard coefficient on BIRD and Spider benchmarks.}
	\label{jaccard}
\end{figure}

As displayed in Table~\ref{accuracy}, Skeleton rewards enhance structural consistency with notable EM gains on Spider, while efficiency rewards optimize the cost-performance trade-off, achieving the highest TEP values on BIRD. Skeleton + Efficiency delivers consistent improvements across benchmarks, achieving peak EM scores on Spider and KaggleDBQA. Most significantly, HES-SQL demonstrates superior effectiveness and the highest EM performance on BIRD, while maintaining excellent token efficiency with TEP scores of 0.83 and 2.04 respectively. This two-stage paradigm consistently outperforms single-stage RL approaches, validating the critical role of supervised initialization in enhancing subsequent RL effectiveness.

Figure~\ref{jaccard} illustrates the cumulative distribution function (CDF) of Jaccard similarity scores across different training strategies on BIRD and Spider benchmarks. The results demonstrate that skeleton-aware training strategies achieve superior structural accuracy: on BIRD benchmark, Skeleton, Skeleton + Efficiency, and HES-SQL show markedly improved distributions with higher median similarities compared to Baseline and Vanilla. HES-SQL exhibits the most favorable distribution, with approximately 80\% of queries achieving Jaccard similarities above 0.6, indicating strong structural fidelity. On Spider benchmark, similar patterns emerge with skeleton-enhanced strategies outperforming baseline approaches. The consistent superiority of skeleton-aware methods across both benchmarks validates the effectiveness of incorporating structural reward signals.

\begin{figure}[t]
	\centering
	\includegraphics[width=0.46\textwidth]{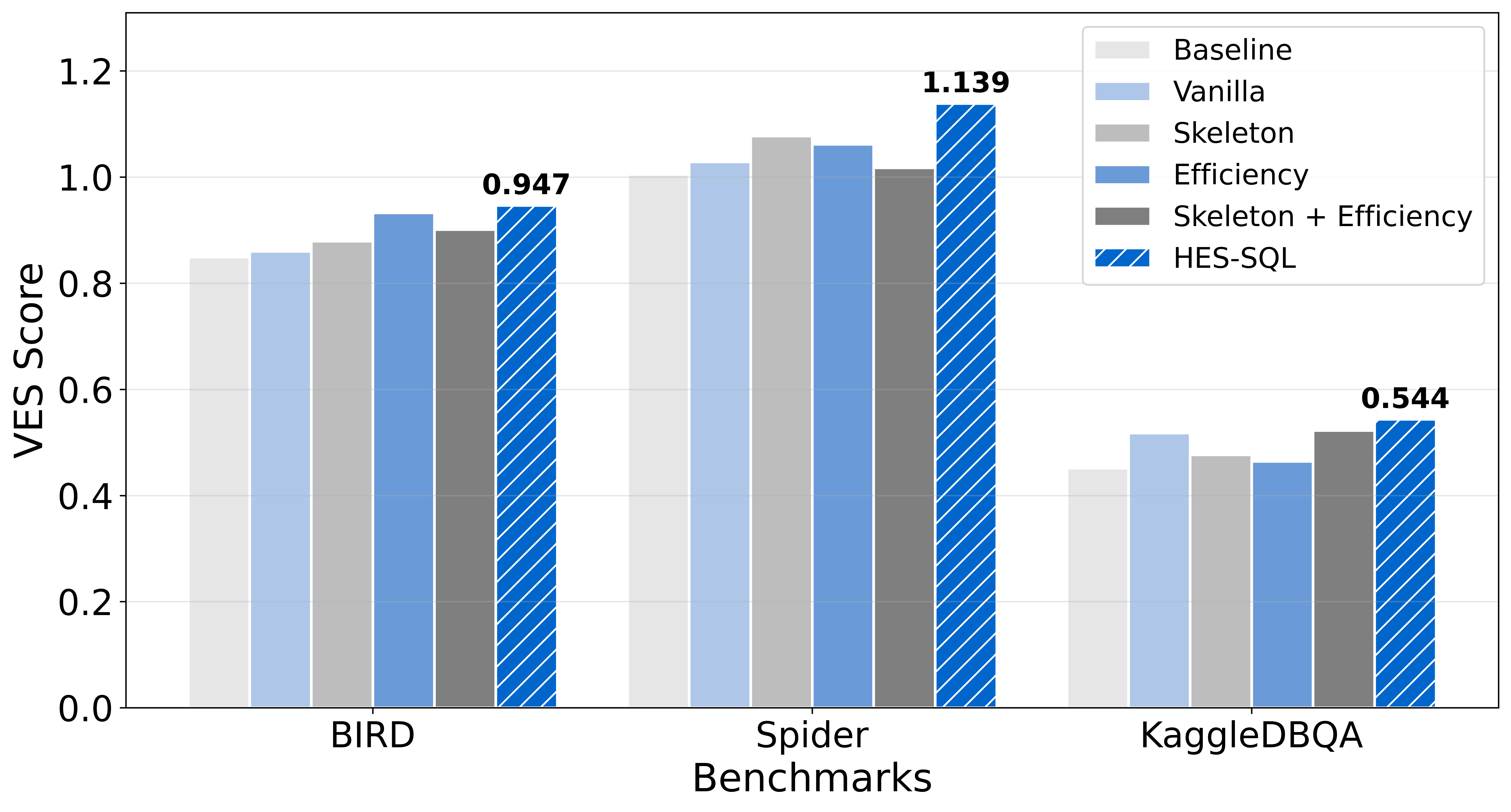}
	\caption{The VES score comparison across models on Text-to-SQL benchmarks.}
	\label{ves_score}
\end{figure}

As shown in Figure~\ref{ves_score}, HES-SQL achieves the highest VES scores across all benchmarks, with notable performance on Spider (1.139) and substantial improvements on BIRD (Approximately 11.42\% over baseline). Efficiency-aware strategies consistently outperform basic approaches, with all achieving VES scores above 1.0 on Spider, indicating generated queries often execute faster than reference solutions while maintaining correctness. The lower VES scores on KaggleDBQA reflect benchmark complexity, though HES-SQL still demonstrates superior efficiency gains. These results validate that incorporating execution efficiency rewards successfully optimizes both correctness and computational performance in Text-to-SQL generation.

\begin{figure}[t]
	\centering
	\includegraphics[width=0.46\textwidth]{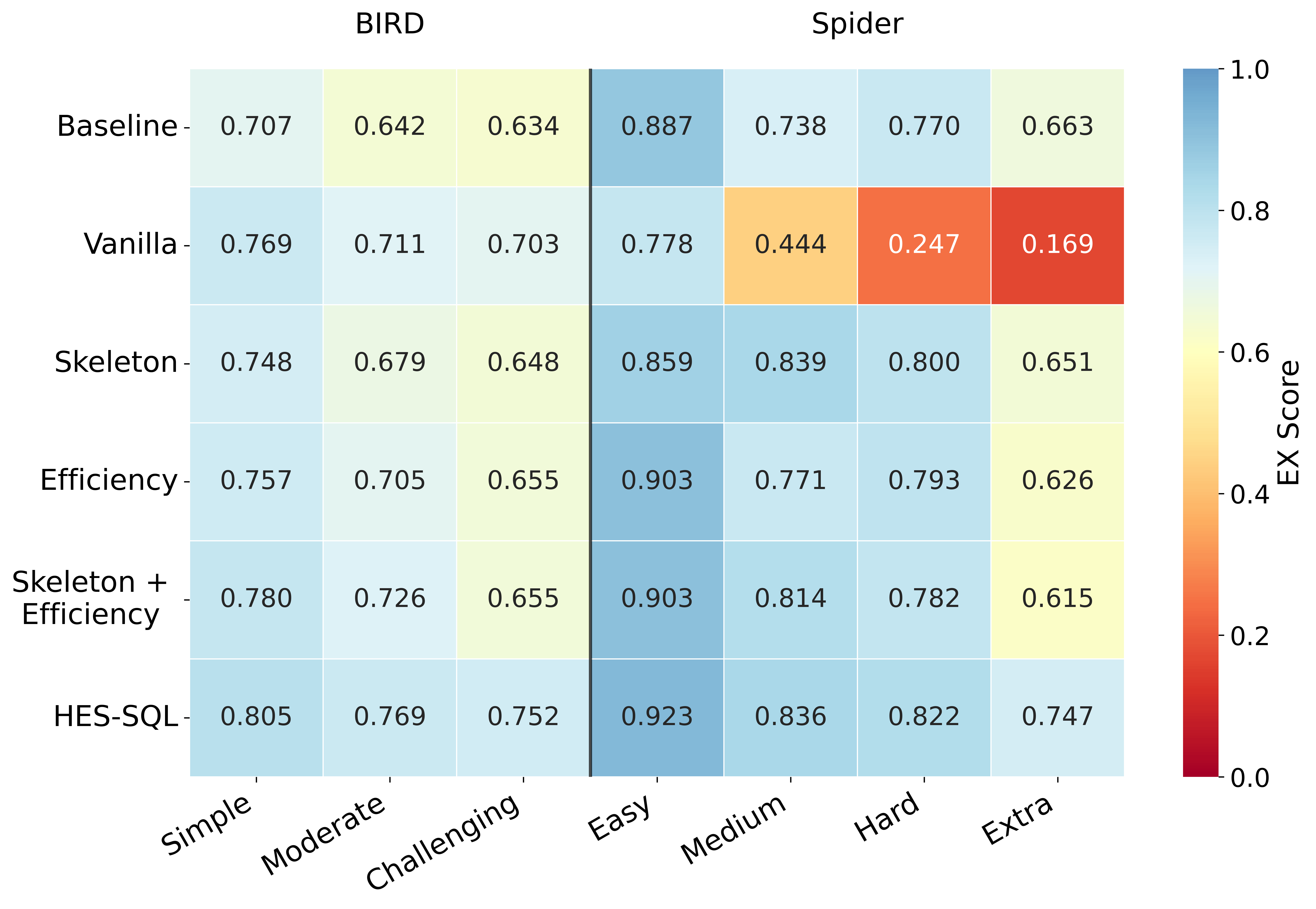}
	\caption{The EX score performance breakdown by difficulty tiers across models on BIRD and Spider benchmarks.}
	\label{heatmap}
\end{figure}

\paragraph{\textbf{Performance Assessment across Difficulty Tiers}}

BIRD is divided into Simple, Moderate, and Challenging categories, while Spider is categorized as Easy, Medium, Hard, and Extra, with accuracy statistics computed for each difficulty level. The color-coded heatmap shows darker blue indicating higher performance and red highlighting poor results (see Figure~\ref{heatmap}). HES-SQL consistently achieves the highest accuracy across all categories, while Vanilla exhibits severe performance degradation on Spider's harder difficulties. BIRD benchmark shows relatively stable performance across strategies, whereas Spider reveals high sensitivity to training methodology with performance gaps up to 0.609. Strategies incorporating comprehensive rewards maintain robust performance across both benchmarks, demonstrating the critical importance of sophisticated reward design.

\begin{figure}[t]
	\centering
	\includegraphics[width=0.46\textwidth]{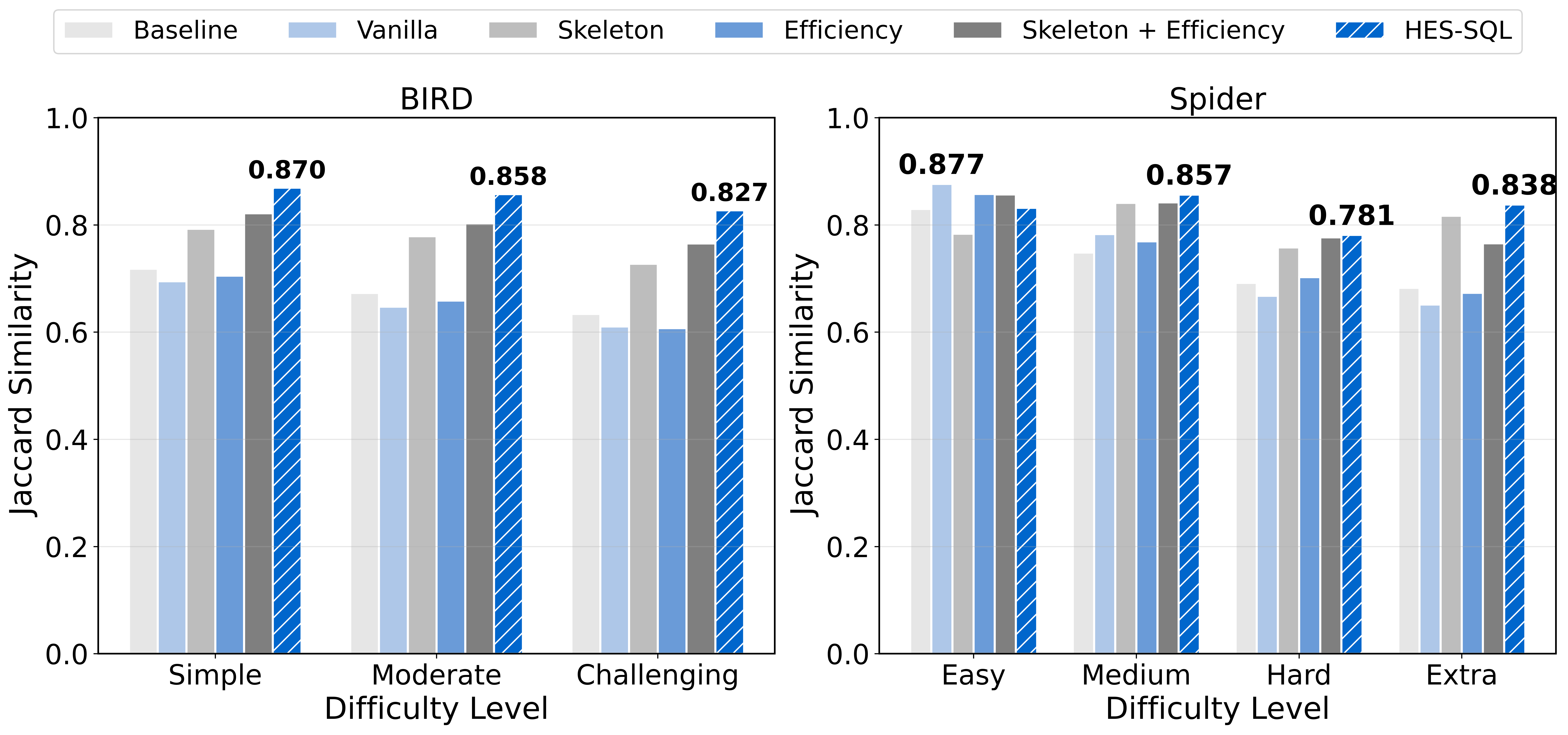}
	\caption{The Jaccard similarity scores across difficulty levels for different models on BIRD and Spider benchmarks}
	\label{jac_tiers}
\end{figure}

In Figure~\ref{jac_tiers}, HES-SQL demonstrates superior structural accuracy, achieving the highest Jaccard similarities across most categories. Skeleton + Efficiency shows strong performance, particularly excelling on BIRD benchmark, while skeleton-enhanced strategies significantly outperform baseline approaches. Notably, Vanilla exhibits inconsistent performance, with modest improvements on Spider Easy but degradation on other categories. The results reveal that BIRD benchmark shows greater sensitivity to skeleton rewards, with Skeleton achieving substantial improvements compared to baseline, while Spider benchmark benefits more from comprehensive reward combinations. The consistent superiority of strategies incorporating skeleton-based structural rewards validates the importance of clause-level and operator-level correspondence in generating syntactically accurate SQL queries.

\begin{figure}[t]
	\centering
	\includegraphics[width=0.46\textwidth]{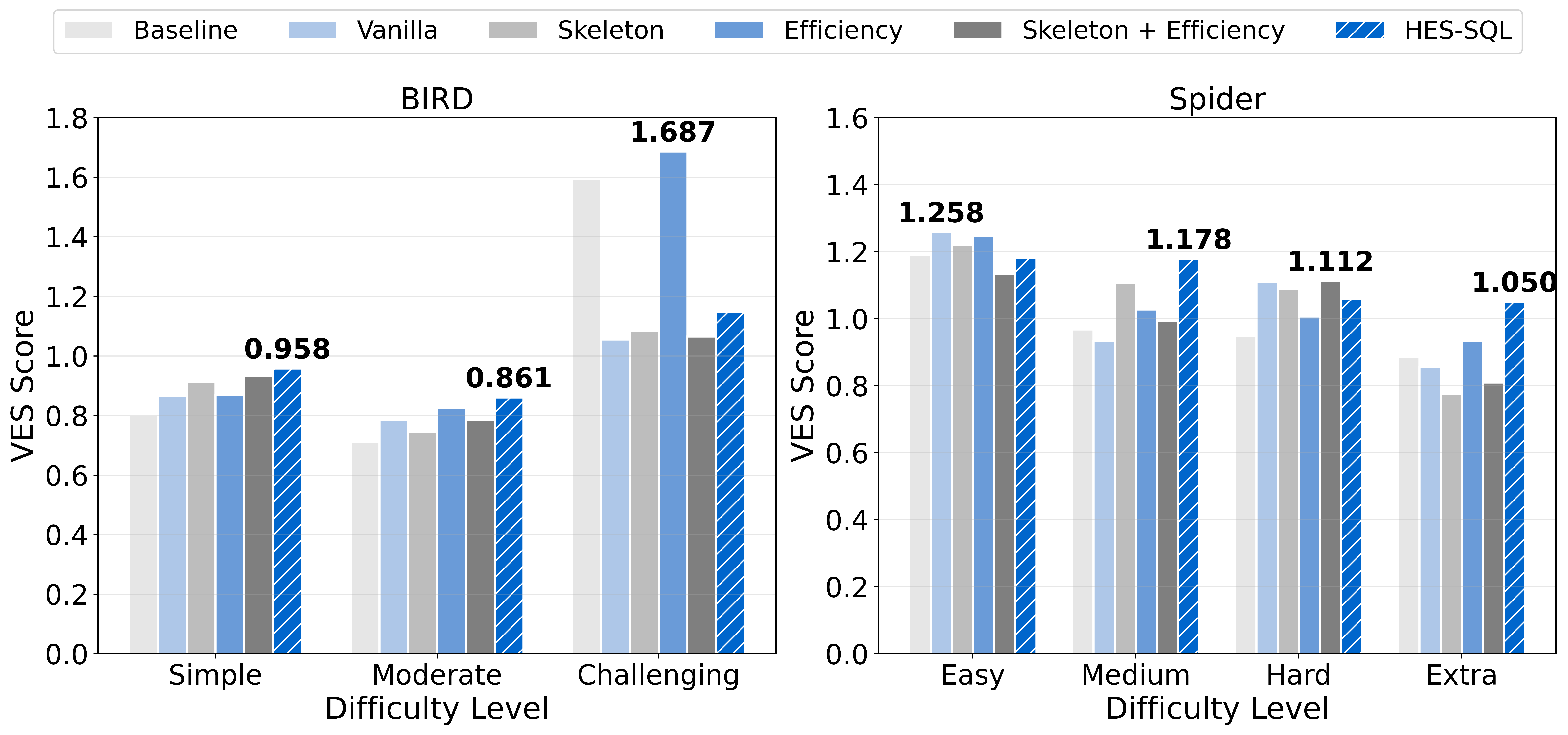}
	\caption{The VES scores across difficulty levels for different models on BIRD and Spider benchmarks.}
	\label{ves_tiers}
\end{figure}

Figure~\ref{ves_tiers} presents VES performance across training strategies using official difficulty classifications. HES-SQL achieves the highest efficiency scores across most categories, demonstrating superior computational performance. Notably, efficiency patterns vary significantly between benchmarks: In BIRD Challenging, generated queries execute substantially faster than reference solutions. On Spider benchmark, all strategies achieve relatively balanced efficiency with scores clustering around 0.8-1.2, suggesting more consistent query complexity. HES-SQL demonstrates the most robust efficiency profile, maintaining strong performance across all difficulty levels without extreme variations. These results indicate that HES-SQL provides the most reliable computational efficiency across diverse Text-to-SQL scenarios.

\begin{figure}[t]
	\centering
	\includegraphics[width=0.46\textwidth]{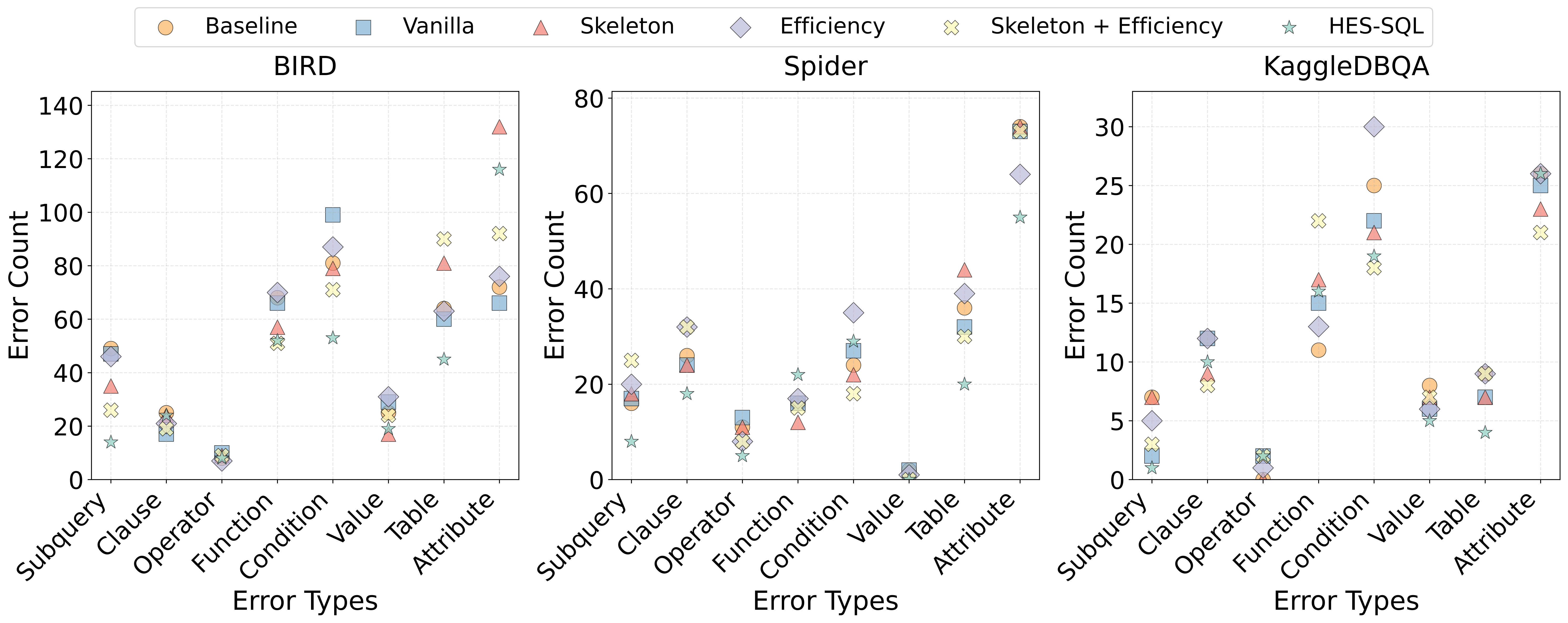}
	\caption{The error type distribution in SQL generation across training strategies on BIRD and Spider benchmarks.}
	\label{error_type_rl}
\end{figure}

\paragraph{\textbf{SQL Error Type Distribution}}

Figure~\ref{error_type_rl} presents the distribution of SQL error types across training strategies on Bird, Spider, and KaggleDBQA benchmarks, categorized into structural (Subquery, Clause, Operator) and semantic (Function, Condition, Value, Table, Attribute) error classes. Examples of SQL error type descriptions are provided in the Appendix. Skeleton-enhanced training demonstrates significant structural error reduction, with Subquery errors decreasing by 29\% on Bird and maintaining effectiveness across all three benchmarks. HES-SQL achieves the most substantial error reduction across benchmarks, notably decreasing Subquery errors by 71\% on Bird, 50\% on Spider, and 86\% on KaggleDBQA, while consistently reducing Condition errors across all datasets. The Skeleton + Efficiency combination shows balanced improvements, reducing critical structural errors while maintaining semantic accuracy. The systematic error reduction pattern confirms that HES-SQL provides superior error mitigation compared to pure reinforcement learning approaches, establishing more robust query generation capabilities across diverse SQL complexity levels and benchmark characteristics.

We observe consistent \emph{relative} efficiency gains across both DBMS engines, with the ranking of variants (by VES) remaining stable between MySQL and SQLite. 
This suggests the latency-aware reward does not overfit a single query engine.

\subsection{Extended Experimental Details}
\label{sec:extended}

\subsubsection{\textbf{Algorithm Workflow}}
\label{app:algorithm}

The overall algorithm workflow can be seen in the right pesudo code block. The RL reward mechanism of HES-SQL employs a multi-component scoring system that validates model responses through format compliance, execution correctness, and schema alignment criteria. The algorithm returns a composite reward score by evaluating SQL query structural similarity, execution results equivalence, performance efficiency, and schema element consistency against ground truth references.

\begin{algorithm}[htbp]
	\SetAlgoLined
	\caption{The RL reward design of HES-SQL}
	\label{alg:nl2sql_rl}
	
	\KwIn{
		$M$: model response, 
		$G$: ground truth SQL,
		$\mathcal{D}$: database collection,
		$\tau_s$: similarity threshold
	}
	\KwOut{$\sigma \in \mathbb{R}$: evaluation score}
	
	\BlankLine
	\tcp{Define scoring components}
	Let $\omega_f, \omega_e, \omega_s, \omega_t \in \mathbb{R}$ be scoring weights\;
	Let $\sigma_f, \sigma_e, \sigma_s, \sigma_t \in \mathbb{R}$ be component scores\;
	
	\BlankLine
	\tcp{Parse response components}
	$(d_{id}, src, mode, think, sql_m) \leftarrow \text{ParseResponse}(M)$\;
	$(\_\_, \_, \_, \_, sql_g) \leftarrow \text{ParseResponse}(G)$\;
	
	\BlankLine
	\tcp{Format validation}
	$\phi(x) \triangleq \text{ThinkingPatternValidator}(x, think)$\;
	\eIf{$\phi(M)$}{
		$\psi(x,y) \triangleq 0.7 \cdot \text{SequenceMatcher}(x,y) + 0.3 \cdot \text{JaccardSim}(x,y)$\;
		$s \leftarrow \psi(\text{SQLSkeleton}(sql_m), \text{SQLSkeleton}(sql_g))$\;
		\eIf{$s \geq \tau_s$}{
			$\sigma_f \leftarrow 1$\;
		}{
			\Return $-2$\;
		}
	}{
		\Return $-2$\;
	}
	
	\BlankLine
	\tcp{Execution validation}
	$DB \leftarrow \mathcal{D}[d_{id}, src]$\;
	$(r_m, t_m) \leftarrow \text{ExecuteSQL}(DB, sql_m)$\;
	$(r_g, t_g) \leftarrow \text{ExecuteSQL}(DB, sql_g)$\;
	
	\eIf{$t_m \leq 30 \land r_m = r_g$}{
		$\sigma_e \leftarrow 2$\;
		$\sigma_t \leftarrow \min(1, \frac{t_g}{t_m})$\;
	}{
		$\sigma_e \leftarrow -2.5$\;
		$\sigma_t \leftarrow 0$\;
		
		\BlankLine
		\tcp{Schema linking validation}
		$S_m \leftarrow \text{ExtractSchemaElements}(sql_m)$\;
		$S_g \leftarrow \text{ExtractSchemaElements}(sql_g)$\;
		
		\eIf{$S_m \subseteq S_g$}{
			$\sigma_s \leftarrow 1.5$\;
		}{
			$\sigma_s \leftarrow 0$\;
		}
	}
	
	\BlankLine
	\Return $\sigma_f + \sigma_e + \sigma_s + \sigma_t$\;
	
\end{algorithm}

\subsubsection{\textbf{Training Data Specifications}}
\label{app:data}

The dataset used in this study is a self-constructed collection designed for the Text-to-SQL task, integrating knowledge from multiple domains. It primarily draws from two corpora: BIRD (9,294 items, averaging 486.84 tokens per question and 17.50 tokens per query) and Spider1Token (5,149 items, averaging 289.91 tokens per question and 15.69 tokens per query). After cleaning, deduplication, and systematic shuffling, the dataset provides diverse linguistic and domain-specific characteristics for robust evaluation.

The dataset combines general knowledge (logical reasoning, multi-task capability, query generation) with domain-specific knowledge (structured query generation, schema understanding) to support enhanced instruction-following and reasoning capabilities. 

For training, we construct separate SFT and RL datasets with different reasoning requirements. The SFT dataset includes high-quality reasoning examples generated by Qwen3 32B—only instances with three consecutive correct answers are retained, using the reasoning process as \texttt{<think>} content to maintain continuity with the base model's thinking patterns. In contrast, the RL dataset enforces only format requirements for reasoning components, mandating \texttt{<think>} tags without constraining content or length, allowing the model to explore diverse reasoning strategies during policy optimization.

\subsubsection{\textbf{Training State Evolution}}

\begin{figure}[t]
	\centering
	\includegraphics[width=0.46\textwidth]{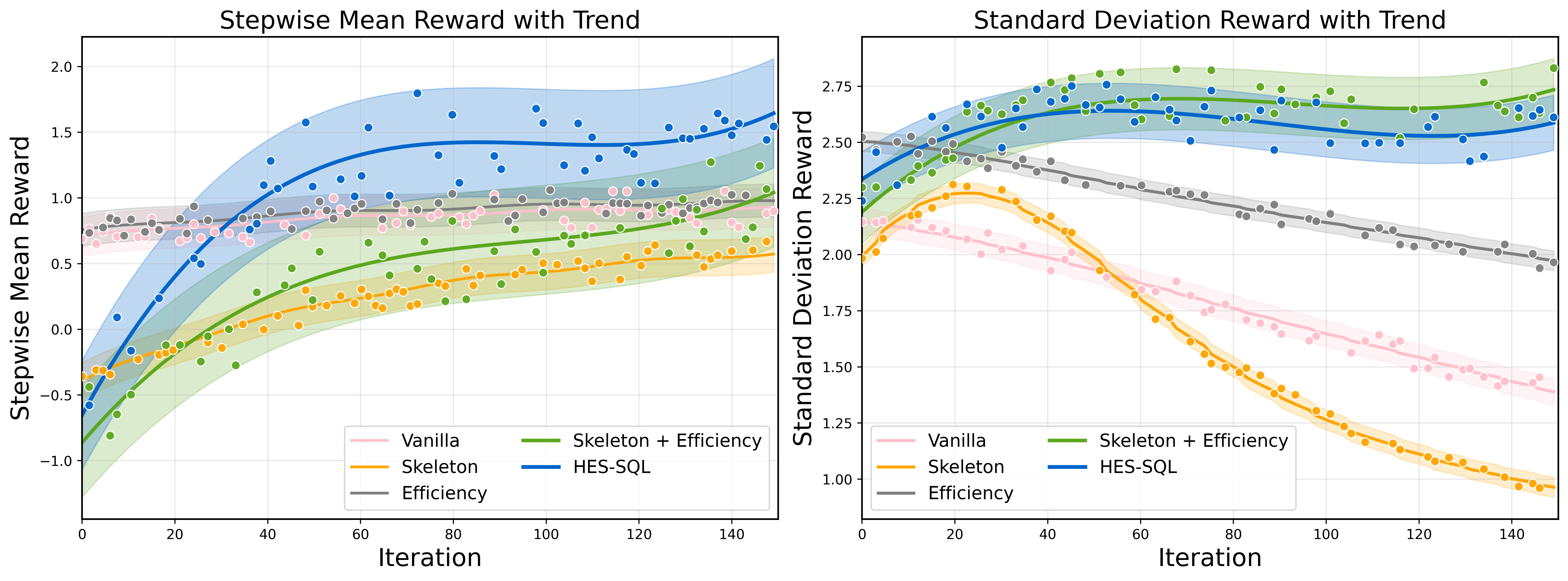}
	\caption{The training reward progression showing mean values and standard deviations across HES-SQL and Skeleton + Efficiency for Text-to-SQL generation.}
	\label{trend}
\end{figure}

Figure~\ref{trend} presents the training dynamics of HES-SQL and Skeleton+Efficiency, which demonstrate superior Text-to-SQL capabilities in our comparative experiments. The step-wise mean reward progression (left panel) reveals that HES-SQL strategy achieves superior performance, reaching the highest mean rewards with remarkable stability after initial supervised fine-tuning, demonstrating the effectiveness of cold-start SFT followed by comprehensive reward-based RL. The standard deviation trends (right panel) indicate HES-SQL maintains the most consistent training behavior, while other strategies show varying degrees of convergence stability. 

\begin{figure}[t]
	\centering
	\includegraphics[width=0.46\textwidth]{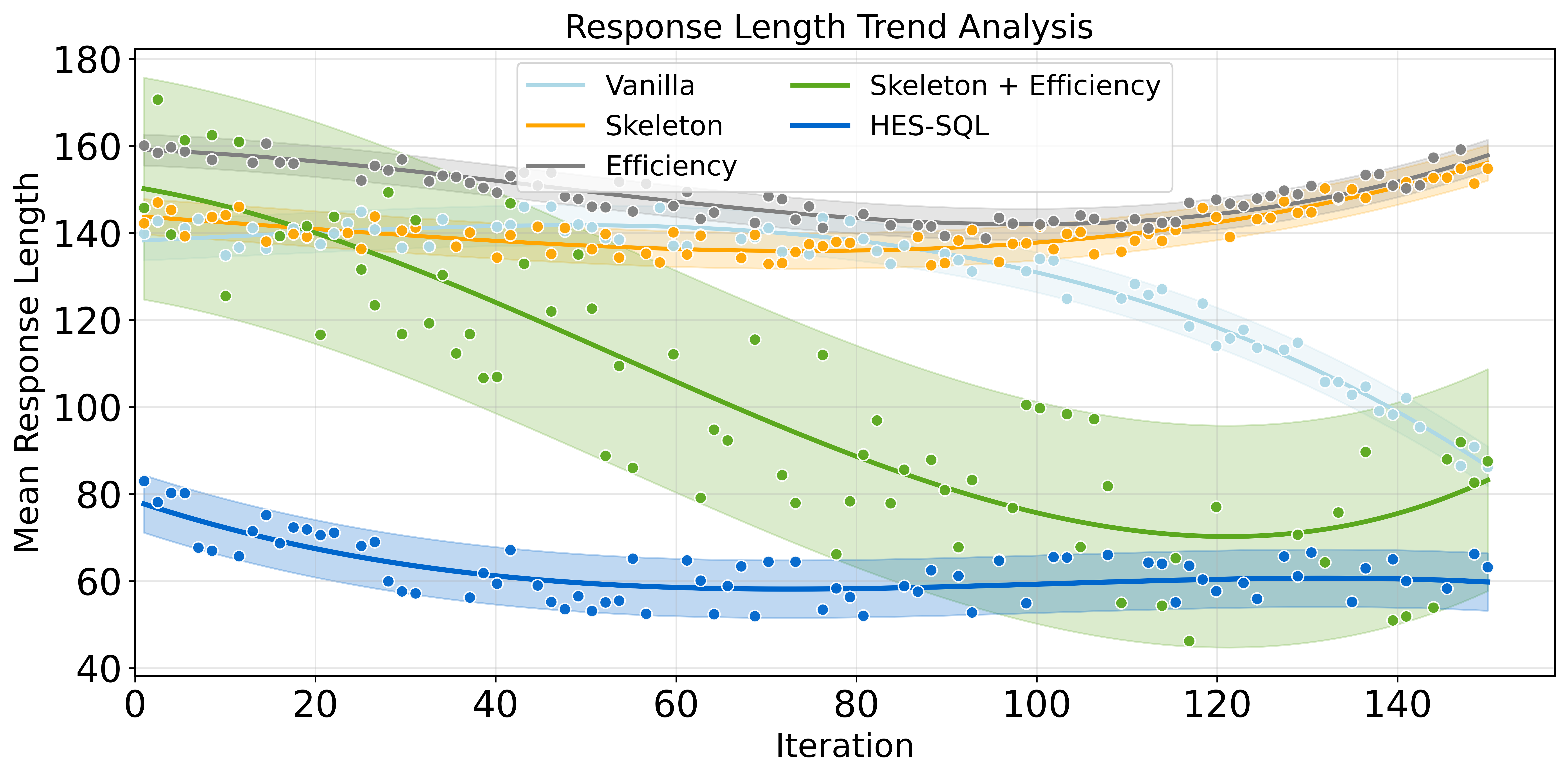}
	\caption{The evolution of mean response length during RL training across HES-SQL and Skeleton + Efficiency.}
	\label{response_length}
\end{figure}

Figure~\ref{response_length} illustrates the evolution of mean response length for HES-SQL and Skeleton+Efficiency. HES-SQL demonstrates superior response length control, maintaining consistently concise outputs throughout training with minimal fluctuation, indicating effective learning of appropriate SQL query brevity. In contrast, Skeleton + Efficiency exhibits problematic response length inflation. This pattern suggests that standard RL approaches struggle with output length control, potentially generating unnecessarily complex or verbose SQL queries. HES-SQL approach's ability to maintain optimal response length while achieving superior reward performance (as shown in Figure~\ref{trend}) demonstrates a crucial advantage of the hybrid training paradigm, producing both high-quality and appropriately concise Text-to-SQL outputs essential for practical deployment.

\subsubsection{\textbf{Evaluation Metrics}}
\label{app:metric}

\paragraph{\textbf{Exact-setmatch Accuracy (EM)}}

The EM \citep{yu2018spider} implements a clause-level evaluation methodology that treats individual clauses as sets, enabling precise comparison between predicted query components and their corresponding ground truth counterparts: 

\begin{equation}\label{eq:em}
	\mathrm{EM} = \frac{1}{N} \sum_{n=1}^{N} \mathds{1}[\mathrm{SetMatch}(Q_n, \hat{Q}_n)], \tag{1}\\
\end{equation}
where $Q_n$ and $\hat{Q}_n$ represent the $n$-th ground truth and predicted SQL queries respectively, $N$ denotes the total query count, and $\mathds{1}[\cdot]$ is the indicator function. Function $\mathrm{SetMatch}(\cdot, \cdot)$ performs clause-level comparison.

\paragraph{\textbf{Execution accuracy (EX)}}

The EX \citep{qin2022survey} employs a result-based evaluation approach that compares the actual execution outcomes of predicted SQL queries with those of reference queries across selected database instances: 

\begin{equation}\label{eq:ex}
	\mathrm{EX} = \frac{1}{N} \sum_{n=1}^{N} \mathds{1}[\mathrm{Exe}(Q_n) = \mathrm{Exe}(\hat{Q}_n)], \tag{2}\\
\end{equation}
where function $\mathrm{Exe}(\cdot)$ denotes the result set of the query.

\paragraph{\textbf{Valid Efficiency Score (VES)}}

The VES \citep{li2023can} serves as a computational efficiency assessment tool, designed to evaluate the runtime performance characteristics of generated SQL query executions:

\begin{equation}\label{eq:ves}
	\mathrm{VES} = \frac{1}{N} \sum_{n=1}^{N} \mathds{1}[\mathrm{Exe}(Q_n) = \mathrm{Exe}(\hat{Q}_n)] \cdot \sqrt{\frac{\mathcal{T}(Q_n)}{\mathcal{T}(\hat{Q}_n)}} \tag{3}
\end{equation}
where $\mathcal{T}(\cdot)$ represents the absolute execution efficiency for SQL queries under specified environmental conditions.

\paragraph{\textbf{Performance Gap Recovered (PGR)}}

We utilize PGR \citep{zhu2025elliesql} to evaluate router performance, capturing the degree to which router $\mathcal{R}$ bridges the performance gap between weak and strong methodological approaches:

\begin{equation}\label{eq:pgr}
	\mathrm{PGR_\mathcal{R}} = \frac{EX_\mathcal{R} - EX_\mathrm{B}}{EX_\mathrm{A} - EX_\mathrm{B}} \tag{4}
\end{equation}

\paragraph{\textbf{Token Elasticity of Performance (TEP)}}

Drawing from economic elasticity principles \citep{li2023llmservedatabaseinterface}, The TEP quantifies the sensitivity of SQL generation pipeline performance to token investment variations. We formalize token consumption as $L = L_{in} + \mu L_{out}$ with average $\bar{L} = L/N$, where $L_{in}$, $L_{out}$, $\mu$, and $N$ denote input tokens, output tokens, completion multiplier, and sample size respectively. TEP can be expressed as:

\begin{equation}\label{eq:tep}
	\mathrm{TEP_\mathcal{M}} = \frac{\Delta{EX_\mathcal{M}} / EX_\mathrm{B}}{\Delta{\bar{L}_\mathcal{M}} / \bar{L}_\mathrm{B}} \tag{5}
\end{equation}
where $\Delta{EX_\mathcal{M}}$ denotes the performance improvement of method $\mathcal{M}$ over baseline $\mathcal{M}_\mathrm{B}$ in execution accuracy, and $\Delta{\bar{L}_\mathcal{M}} = \bar{L}_\mathcal{M} - \bar{L}_\mathrm{B}$ represents the corresponding incremental token consumption.

\subsubsection{\textbf{Sample for SQL Error Types}}
\label{app:error}

Typical error types observed in benchmarks for SQL query generation are illustrated with specific examples and reference answers (Gold) in Table~\ref{tab:error}, where bold text highlights the characteristic details of each error type.

\subsubsection{\textbf{General Capability Retention}}

The evaluation encompasses three general capability benchmarks: IFEval assesses instruction-following ability \citep{zhou2023instructionfollowingevaluationlargelanguage}, GPQA evaluates scientific reasoning and knowledge comprehension \citep{rein2023gpqagraduatelevelgoogleproofqa}, and MATH-500 measures mathematical problem-solving skills \citep{hendrycks2021measuringmathematicalproblemsolving}. As shown in Table~\ref{tab:general}, HES-SQL demonstrates effective preservation of general capabilities while enhancing Text-to-SQL performance, with IFEval scores improving from 76.71 to 80.04, indicating that the specialized training does not compromise and may even enhance the model's broader competencies.

\begin{table}[htbp]
	\centering
	\caption{General performance comparison of HES-SQL and the Baseline.}
	\label{tab:general}
	\begin{tabular}{lccc}
		\toprule
		Model & IFEval & GPQA & MATH-500 \\
		\midrule
		Baseline & 76.71 & 53.36 & 86.2 \\
		HES-SQL & \textbf{80.04} & 43.37 & 81.4 \\
		\bottomrule
	\end{tabular}
\end{table}

\begin{table*}[htbp]
	\centering
	\caption{The examples of SQL query error types and their corresponding failure cases.}
	\label{tab:error}
	\renewcommand{\arraystretch}{1.2}
	\begin{tabular}{|p{4cm}|p{3cm}|p{5cm}|p{5cm}|}
		\toprule
		Error type & Query & Gold & Model output \\
		\midrule
		Condition-related error & List all the elements of atoms that can not bond with any other atoms. & SELECT DISTINCT T.element FROM atom AS T WHERE T.element NOT IN (SELECT DISTINCT T1.element FROM atom AS T1 INNER JOIN connected AS T2 ON T1.atom\_id = T2.atom\_id) & SELECT atom\_id FROM atom WHERE atom\_id NOT IN (SELECT atom\_id FROM connected) AND \textbf{element IN ('cl', 'c', 'h', 'o', 's', 'n', 'p', 'na', 'br', 'f', 'i', 'sn', 'pb', 'te', 'ca')} \\
		\hline
		Operator-related error & How many patients born after $\frac{1980}{1}/1$ have an abnormal fibrinogen level? & SELECT COUNT(DISTINCT T1.ID) FROM Patient AS T1 INNER JOIN Laboratory AS T2 ON T1.ID $=$ T2.ID WHERE T2.FG $<=$ 150 OR T2.FG $>=$ 450 AND T1.Birthday $>$ '1980-01-01' & SELECT COUNT(DISTINCT T1.ID) FROM Patient AS T1 JOIN Laboratory AS T2 ON T1.ID $=$ T2.ID WHERE T1.Birthday $>$ '1980-01-01' AND (T2.FG \textbf{$<$} 150 OR T2.FG \textbf{$>$} 450) \\
		\hline
		Function-related error & How many British drivers were born after 1980? & SELECT COUNT(driverId) FROM drivers WHERE nationality = 'British' AND STRFTIME('\%Y', dob) $>$ '1980' & SELECT COUNT(*) FROM drivers WHERE nationality $=$ 'British' AND \textbf{dob $>$ '1980-01-01'} \\
		\hline
		Value-related error & From 2000 to 2005, what percentage of drivers who were born before 1985 and the lap numbers were over 50? & SELECT CAST(SUM(IIF(STRFTIME('\%Y', T3.dob) $<$ '1985' AND T1.laps $>$ 50, 1, 0)) AS REAL) * 100 / COUNT(*) FROM results AS T1 INNER JOIN races AS T2 on T1.raceId $=$ T2.raceId INNER JOIN drivers AS T3 on T1.driverId $=$ T3.driverId WHERE T2.year BETWEEN 2000 AND 2005 & SELECT CAST(SUM(CASE WHEN strftime('\%Y', drivers.dob) \textbf{$<$ 1985} AND results.laps $>$ 50 THEN 1 ELSE 0 END) AS REAL) * 100 / COUNT(results.driverId) AS percentage FROM results JOIN drivers ON results.driverId $=$ drivers.driverId JOIN races ON results.raceId $=$ races.raceId WHERE races.year BETWEEN 2000 AND 2005 \\
		\hline
		Table-related error & How many driver participated in race ID number 18? & SELECT COUNT(driverId) FROM driverStandings WHERE raceId $=$ 18 & SELECT COUNT(DISTINCT T1.driverId) \textbf{FROM results} AS T1 WHERE T1.raceId $=$ 18 \\
		\hline
		Attribute-related error & Indicate the full names of the top 10 players with the highest number of penalties. & SELECT t2.player\_name FROM Player\_Attributes AS t1 INNER JOIN Player AS t2 ON t1.id $=$ t2.id ORDER BY t1.penalties DESC LIMIT 10 & SELECT T2.player\_name FROM Player\_Attributes AS T1 JOIN Player AS T2 ON \textbf{T1.player\_api\_id $=$ T2.player\_api\_id} WHERE T1.penalties IS NOT NULL ORDER BY T1.penalties DESC LIMIT 10 \\
		\hline
		Subquery-related error & List the players' api id who had the highest above average overall ratings in 2010 & SELECT player\_api\_id FROM Player\_Attributes WHERE SUBSTR(`date`, 1, 4) $=$ '2010' ORDER BY overall\_rating DESC LIMIT 1 & SELECT player\_api\_id FROM Player\_Attributes WHERE substr(date,1,4) $=$ '2010' AND overall\_rating = \textbf{(SELECT MAX(overall\_rating) FROM Player\_Attributes WHERE substr(date,1,4) $=$ '2010')} \\
		\hline
		Clause-related error & Which are the top five best selling products? Please state the full name of them & SELECT T2.Description FROM transactions\_1k AS T1 INNER JOIN products AS T2 ON T1.ProductID $=$ T2.ProductID ORDER BY T1.Amount DESC LIMIT 5 & SELECT T2.Description FROM transactions\_1k AS T1 JOIN products AS T2 ON T1.ProductID $=$ T2.ProductID \textbf{GROUP BY T1.ProductID ORDER BY SUM(T1.Amount)} DESC LIMIT 5 \\
		\bottomrule
	\end{tabular}
\end{table*}

\section{Related Work}

Recent advances in Text-to-SQL have witnessed substantial progress driven by large-scale language models and novel training strategies \citep{liu2025xiyansqlnovelmultigeneratorframework, li2025omnisqlsynthesizinghighqualitytexttosql}. Building upon the foundational challenges and methodological developments outlined in the previous section, we now examine how our work relates to and extends beyond existing approaches.

Our work builds upon these advances with several key innovations. We integrate self-distillation with a novel RL fine-tuning scheme using composite rewards to address both semantic accuracy and execution efficiency. By incorporating skeleton-based structural rewards and latency-based performance rewards into group-based policy optimization, HES-SQL simultaneously addresses reward sparsity and efficient query execution. To the best of our knowledge, this is the first Text-to-SQL framework that explicitly optimizes query latency as part of the training objective. Moreover, progressive self-distillation maintains the model's chain-of-thought reasoning capability, distinguishing our approach from prior RL-only methods. These contributions enable HES-SQL to effectively balance correctness and efficiency in Text-to-SQL generation.

\section{Conclusion}

HES-SQL addresses reasoning complexity and execution efficiency in Text-to-SQL generation through a hybrid training strategy combining SFT with GRPO. The skeleton-based reward filtering improves accuracy by 10.42\%–18.06\% over baselines, while the latency-aware scoring system consistently produces SQL queries that execute faster than reference solutions. Latency is measured on MySQL 8.0 and SQLite 3.42 under controlled single-user conditions, ensuring reproducible efficiency comparisons across two representative DBMS engines. 

Overall, HES-SQL achieves execution accuracies of 79.14\% and 54.9\% on BIRD and KaggleDBQA, ranking among top-performing open-source systems. The self-distillation step introduces only modest additional computation—scaling with checkpoint reuse—yet preserves hybrid reasoning capabilities and prevents degradation during fine-tuning. The proposed framework effectively balances semantic accuracy with execution performance, establishing a new paradigm for Text-to-SQL systems that accounts for real-world efficiency constraints. We have deployed this method as part of the CEM Copilot application at Saudi~STA sites, demonstrating its practicality in an industry setting.

\paragraph*{\textbf{Deployment lessons.}}
Our production deployment within the CEM Copilot application at Saudi~STA sites surfaced several pragmatic findings. First, efficiency-aware training reduces tail latency (e.g., long-running joins) and thus stabilizes end-to-end user experience. Second, skeleton filtering cuts down DBMS error spikes (e.g., missing columns or ill-formed subqueries), lowering incident rates without hand-engineered guards. Third, thinking-mode control helps enforce length/complexity budgets in conversational settings, which is crucial for predictable cost and responsiveness.

\paragraph*{\textbf{Limitations and threats to validity.}}
HES-SQL currently targets single-user execution on MySQL and SQLite; while relative VES rankings remain consistent across engines, validation on additional DBMSes and concurrent workloads is needed. Our latency reward uses wall-clock runtime, which is robust but may be affected by caching effects and schema-specific physical designs. The self-distillation pipeline filters model-generated rationales using exact-match criteria, which, while conservative, may exclude semantically equivalent solutions. Finally, some benchmarks use simplified schemas; real enterprise deployments involving access controls, materialized views, or user-defined functions may introduce failure modes not captured here.

\paragraph*{\textbf{Broader impact.}}
By explicitly accounting for execution efficiency during training, HES-SQL pushes NL2SQL evaluation toward a more systems-aware objective that better reflects production constraints. This shift encourages research on the accuracy–latency frontier rather than accuracy alone, and it opens a path for principled collaboration between learned parsers and database optimizers.

\section{Future Work}
We outline several directions to advance the accuracy–efficiency frontier and strengthen real-world readiness:
\begin{itemize}[leftmargin=1.2em]
	\item \emph{Optimizer-in-the-loop rewards.} Replace or complement wall-clock latency with plan-level signals (estimated cost, cardinality, join order) and plan-shape regularizers. A learned critic over query plans could provide dense, engine-agnostic guidance while reducing runtime variance.
	
	\item \emph{Cross-engine and multi-tenant generalization.} Extend training to additional DBMS backends (PostgreSQL, commercial engines, cloud warehouses) and evaluate under concurrent workloads. Domain randomization over indexes, statistics, and hardware profiles could improve robustness.
	
	\item \emph{Plan-space exploration and repair.} Augment GRPO with structured search over logical/physical plan neighborhoods (predicate pushdown, join reordering). When execution fails, use counterfactual edits to repair queries while preserving semantics.
	
	\item \emph{Schema evolution and drift.} Introduce continual learning to adapt to schema changes (renames, splits, new columns) with minimal forgetting, using weak supervision from failed executions and user confirmations.
	
	\item \emph{Interactive and trustworthy NL2SQL.} Calibrate uncertainty to trigger clarification questions or safer fallbacks. Combine confidence scoring with selective chain-of-thought disclosure to balance transparency and privacy.
	
	\item \emph{Richer efficiency objectives.} Extend beyond latency to multi-objective rewards incorporating memory footprint, spill events, and quota adherence, enabling SLO compliance in shared clusters and pay-as-you-go environments.
	
	\item \emph{Evaluation and benchmarking.} Complement existing benchmarks with scenario-driven suites including evolving schemas, noisy NL, and explicit cost constraints, encouraging reporting of accuracy, efficiency, and stability metrics.
\end{itemize}

\section*{Acknowledgments}
	This work was supported by the AI Service Capability 26.0 Offering Business Plan (OBP) Development Project of Huawei Technologies Co., Ltd. (Project No. 9463674).

\clearpage
\bibliography{custom}

\end{document}